\documentclass[prc,twocolumn,showpacs,preprintnumbers,amsmath,amssymb]{revtex4-1}
\usepackage[utf8]{inputenc}
\usepackage{epsfig}
\usepackage{graphicx}
\usepackage{tabularx}
\usepackage{xcolor}
\newcolumntype{b}{>{\hsize=1.3\hsize}X}
\newcolumntype{s}{>{\hsize=.3\hsize}X}
\usepackage{bm}
\usepackage{array}
\usepackage{hyperref}
\allowdisplaybreaks

\begin{document}
\title{Electroproduction of medium- and heavy-mass hypernuclei}
\author{P. Byd\v{z}ovsk\'y$^{1}$, D. Denisova$^{1}$, F. Knapp$^{2}$, 
and P. Vesel\'y$^{1}$}
\affiliation{$^{1}$Nuclear Physics Institute, ASCR, 
25068 \v{R}e\v{z}/Prague, Czech Republic \\
$^{2}$Institute of Particle and Nuclear Physics, Faculty 
of Mathematics and Physics,\\ Charles University, 
V Hole\v{s}ovi\v{c}k\'{a}ch 2, 18000 Prague, Czech Republic }

\date{\today }
%
%
\begin{abstract}
The DWIA formalism for computing the cross sections in electroproduction 
of hypernuclei used before for lighter ($p$-shell and $sd$-shell) 
systems is utilized in studying  electroproduction 
on heavier targets such as $^{52}$Cr, and $^{208}$Pb. First, the  
effects from kaon distortion and kinematics are re-examined and then 
results for the heavy targets are discussed in view of various forms 
of the $\Lambda N$ effective interaction, the elementary amplitude, 
and kinematics. Apparent sensitivity of the hypernucleus excitation 
spectra to various forms of the $\Lambda N$ effective interaction is shown.  
Predictions of the excitation spectrum in electroproduction of 
$^{208}_{~~\Lambda}$Tl are also given in kinematics of the experiment 
E12-20-013 in preparation at Jefferson Lab.
\end{abstract}
\maketitle
%
%
\section{Introduction}
Investigation of hypernucleus properties provides significant 
information about the baryon-baryon interaction with a non zero 
strangeness. Processes involving production of hypernuclei play 
important role, as they allow one to study effective forms of the 
interaction. The measured excitation spectra and cross sections give 
information about details of the interaction. To this end, one should 
have a good understanding of the reaction mechanism to be able to draw 
firm conclusions from a data analysis.
 
The hypernuclei have been already studied for several decades in various 
reactions~\cite{GHM2016,HT2006}. The quality of obtained data depends on 
the reaction where the best precision was achieved in the measurement of 
$(\pi^+, K^+\gamma)$~\cite{HT2006}. These high-quality $\gamma$-ray data 
for $p$-shell hypernuclei allowed performing a detailed analysis of the 
spin-dependent part of the effective hyperon-nucleon (YNG) 
interaction~\cite{JohnM}. 
The hypernuclei were also studied in the $(\pi^+, K^+)$ 
reaction~\cite{Pile,HT2006} providing mainly information about the 
binding energies of $\Lambda$ in various shells and the cross sections 
for production of the ground state. 
Reaction spectra with a better energy resolution than in 
$(\pi^+, K^+)$ were obtained from the $(e, e^\prime K^+)$ reaction  
measured for the first time at Jefferson Lab~\cite{HallCexp}. 
Data from the successive measurements of electroproduction of 
$p$-shell hypernuclei performed at Jefferson Lab then allowed studying  
the YNG interaction~\cite{archiv}. 
Heavier hypernuclei beyond the $p$-shell were investigated in the past 
utilizing mainly the $(\pi^+, K^+)$ reaction~\cite{Pile}. However, new 
experiments for electroproduction of hypernuclei on the medium-mass,  
$^{40}$Ca and $^{48}$Ca (E12-15-008~\cite{E12-15-008}), and heavy, 
$^{208}$Pb (E12-20-013~\cite{E12-20-013}), targets are planned at 
Jefferson Lab. 

Electroproduction of $\Lambda$ hypernuclei not only provides a test 
ground for various effective YNG interactions but it also allows 
studying the reaction mechanism and the elementary amplitudes at 
very small kaon angles. 
The theoretical photoproduction excitation spectra obtained in the 
distorted-wave impulse approximation (DWIA) for the $p$-shell and 
medium-mass hypernuclei were discussed by Motoba {\it et al.} in 
Refs.~\cite{Motoba2010} and \cite{NPA881}. It was shown 
that the spectra for typical medium-mass hypernuclei such as 
$^{28}_{~\Lambda}$Al and $^{40}_{~\Lambda}$K provide interesting 
opportunities of spectroscopic study beyond the $p$ shell. 
Another analysis of the cross sections in photoproduction of light 
($^{12}_{~\Lambda}$B) as well as very heavy ($^{208}_{~~\Lambda}$Tl) 
hypernuclei was performed in Ref.~\cite{MotobaJPS} showing the wide 
applicability of the ($e,e^\prime K^+$) reaction. 
In our previous analysis~\cite{HYP2018,fermi,structure} of hypernucleus 
electroproduction we studied the kinematical and Fermi motion effects 
in the case of $p$-shell hypernuclei ($^{12}_{~\Lambda}$B, 
$^{16}_{~\Lambda}$N, $^9_\Lambda$Li)~\cite{fermi} and verified that 
the nucleus and hypernucleus structure of medium-mass systems 
($^{28}_{~\Lambda}$Al, $^{40}_{~\Lambda}$K) can be realistically 
described in the Tamm-Dancoff approach~\cite{structure}. 
Here we continue our study, focusing on the medium- and heavy-mass 
hypernuclei and providing predictions for the experiments E12-15-008 
and E12-20-013.  

The paper is organized as follows: in Sec. II we give a brief overview 
of basic formalism. More details can be found in Refs.~\cite{fermi} 
and \cite{structure}. In Section III, the effects from kaon distortion, 
kinematics, and Fermi motion are discussed in the case of heavy 
hypernuclei.  
Here we also elaborate the optimum on-shell approximation introduced 
in Ref.~\cite{fermi}. The excitation spectra in electroproduction of 
medium- and heavy-mass hyperncuclei, $^{40}_{~\Lambda}$K, 
$^{48}_{~\Lambda}$K, $^{52}_{~\Lambda}$V, and $^{208}_{~~\Lambda}$Tl, 
are discussed in Sec.~IV.   
The summary and conclusions from our study are given in Sec.~V.  
  
%
\section{Basic formalism}

The cross section in electroproduction of hypernuclei is calculated in 
the impulse approximation (IA) considering the optimal factorization 
approximation where the amplitude for elementary production is evaluated 
for an effective proton momentum~\cite{fermi}. 
The general form of the two-component 
elementary amplitude, constructed in Ref.~\cite{fermi}, allows for an 
arbitrary value of this momentum. In the previous analysis we had suggested 
an optimum value of the effective proton momentum which fulfills the 
necessary kinematical conditions and for which the elementary amplitude is on-shell. 
In the many-particle matrix element, the photon plane wave and kaon 
distorted wave are decomposed into partial waves ($LM$) and the target 
proton and final $\Lambda$ are supposed to occupy the single-particle 
states $\alpha$ ($=n,l,j$) and $\alpha^\prime$, respectively.
The reduced amplitude, used to compute the cross sections, then reads
\begin{eqnarray}
A^\lambda_{Jm} =&&\frac{1}{[J]}\,\sum_{S\eta}\,{\cal F}^S_{\lambda\eta}\,
\sum_{LM}\;{\sf C}^{Jm}_{LMS\eta}\sum_{\alpha'\alpha}
{\cal R}^{LM}_{\alpha'\alpha}\;{\cal H}^{LSJ}_{l'j'lj}\nonumber\\
&&\times(\,\Phi_H\,||\,[b_{\alpha'}^+\otimes a_\alpha]^J\,||\;\Phi_A\,)\,,
\label{amplitude-3}
\end{eqnarray}
where ${\cal F}^S_{\lambda\eta}$ are spherical elementary amplitudes with 
a baryon spin-flip $S=0, 1$ and photon helicity $\lambda= \pm 1, 0$. 
The nucleus($\Phi_A$)-hypernucleus($\Phi_H$) transition is 
described by the radial integral ${\cal R}^{LM}_{\alpha'\alpha}$ and the 
one-body density matrix element (OBDME), where $a_\alpha$ and 
$b^+_{\alpha^\prime}$ are the proton annihilation and $\Lambda$ particle 
creation operators. The term ${\cal H}^{LSJ}_{l'j'lj}$ includes the 
Racah algebra and ${\sf C}^{Jm}_{LMS\eta}$ is the Clebsch-Gordan 
coefficient. Derivation of this formula and expressions for the cross 
sections can be found in Ref.~\cite{fermi}.

The radial integrals are explicitly  
\begin{equation}
{\cal R}^{LM}_{\alpha^\prime\alpha} = \int_0^\infty \!\!d\xi\,\xi^2\,
{R}_{\alpha^\prime}(\xi)^*\,F_{LM}(\Delta B\xi)\,{R}_{\alpha}(\xi)\,, 
\label{radial}
\end{equation}
where $R_{\alpha^\prime}$ and $R_\alpha$ are the $\Lambda$ and proton 
radial wave functions, respectively, and $F_{LM}$ comes from the 
partial-wave decomposition of the photon and kaon waves, see  
appendix C in Ref.~\cite{fermi}. This function depends on the momentum 
transfer $\Delta=|\vec{P_\gamma}-\vec{P}_K|$ and parameter 
$B= (A-1)/(A-1+m_\Lambda/m_p)$. Note that in the case without 
a kaon distortion in PWIA the function $F_{LM}$ corresponds to the 
spherical Bessel function.

Kaon distortion is in our approach included by means of the eikonal 
approximation with the first-order optical potential. The kaon 
distorted wave then reads as~\cite{DThesis}
\begin{equation}
\chi^+_{\rm K}(\vec{r}\,)= 
{\sf exp}\left[-b\,\sigma^{tot}_{\rm KN}(s)\!\left(\!\frac{}{}1-i\alpha(s)\right)\!\!
\int_0^\infty\!\!\!dt\,\rho(\vec{r}+\hat{p}\,t) \right],
\label{kaonwave}
\end{equation}
where $\sigma^{tot}_{\rm KN}$ is the kaon-nucleon ($KN$) total cross 
section and 
$\alpha(s)= {\rm Re} f_{\rm KN}(s,0)/{\rm Im} f_{\rm KN}(s,0)$,  
with the $KN$ amplitude at zero angle and invariant energy squared $s$.  
The parameter $b$ stands for a kinematical factor~\cite{DThesis} and the 
integrand  $\rho(\vec{r}+\hat{p}\,t)$ describes the nucleon density 
where $\hat{p}$ is a unit vector in the direction of the relative 
kaon-hypernucleus momentum.  

The nucleus and hypernucleus structure included in the OBDME, 
$(\,\Phi_H\,||\,[b_{\alpha'}^+\otimes a_\alpha]^J\,||\;\Phi_A\,)$, and 
the radial single-particle wave functions $R_\alpha$ are calculated using 
a many-particle formalism with specific $NN$ and $YN$ effective interactions.
In the present analysis we use the Hartree-Fock (HF) and Tamm-Dancoff 
(TD$_{\Lambda}$) approaches \cite{Ves1,structure}.  
In Refs. \cite{Ves1,structure} the HF equation was solved for the hypernuclear 
Hamiltonian including two-body ($NN$, $N\Lambda$), and three-body 
($NNN$) interactions. In this paper we do not include the three-body $NNN$  
interaction term. Instead, to mimic the many-nucleon interactions in 
the nuclear Hamiltonian, we complement the $NN$ interaction with 
a phenomenological density-dependent (DD) term \cite{DDterm}
\begin{equation}
v_{\rho} = \frac{C_{\rho}}{6} (1+P_{\sigma}) \rho\left( \frac{\vec{r}_1 +
\vec{r}_2}{2} \right) \delta(\vec{r}_1-\vec{r}_2),
\label{DD}
\end{equation}
where $C_{\rho}$ is a coupling constant which plays role of a free 
parameter, $P_{\sigma}$ is the operator of the spin exchange, and $\rho$ 
is the nucleon density of the calculated nucleus. 

Applying the HF method for description of $^{40}$Ca, $^{48}$Ca, 
$^{52}$Cr, and $^{208}$Pb we obtain the proton and neutron single-particle 
states. The $\Lambda$ single-particle states are obtained by solving 
the additional Schr\"odinger-like equation (see Eq. (4) in \cite{Ves1}).   
In the case of $^{52}$Cr, which consists of 24 protons and 28 neutrons, 
we solve the HF equations by using the approximation of partial occupation 
of the proton valence $0f_{7/2}$ level by 4 instead of 8 protons. 

The hypernuclei $^{40}_{~\Lambda}$K, $^{48}_{~\Lambda}$K, 
$^{52}_{~\Lambda}$V, and $^{208}_{~~\Lambda}$Tl are described within 
TD$_{\Lambda}$ approach \cite{Ves1,structure} in which we diagonalize 
the hypernuclear Hamiltonian in the space of $\Lambda$-particle 
proton-hole excitations on top of the HF states of $^{40}$Ca, 
$^{48}$Ca, $^{52}$Cr, and $^{208}$Pb, respectively.  

The HF + TD$_{\Lambda}$ calculations are performed within a harmonic 
oscillator (HO) basis, which should be large enough in order to reach 
convergence with respect to a HO frequency. 
It is sufficient to use the space up to the $N_{\rm{max}}=12$ major shell.  

The nucleon-nucleon interaction is described by the 
Daejeon 16~\cite{Dae16-ref,Dae16-code} $NN$ interaction complemented with 
the phenomenological DD term (\ref{DD}) which we denote as D16+DDT. 
The Daejeon 16 interaction itself is derived from the $NN$ component of 
the chiral N3LO potential~\cite{N3LO}, subsequently softened by a SRG 
transformation \cite{SRG} with flow parameter $\lambda$ = 1.5 fm$^{-1}$. 
The phenomenological DD term mimic the effects of the 3-body and 
higher-body nucleon interactions and its coupling constant $C_{\rho}$ 
was tuned for each nucleus separately in order to obtain a realistic 
description of nuclear radii and nucleon single-particle energies. 
The $N\Lambda$ interaction is described using various effective G-matrix 
potentials based on the $YN$ interactions Nijmegen F~\cite{YNG}, 
J\"ulich A~\cite{YNG}, and chiral LO~\cite{YNLO}. The G-matrix derived 
from the Nijmegen F, and J\"ulich A interactions is parametrized as 
a sum of Gaussian-like terms 
 \begin{equation}
    V_{\Lambda N} = \sum^{3}_{i=1} (a_i + b_i k_{\rm{F}} + 
    c_i k^2_{\rm{F}} )\,\rm{exp}(-r^2/\beta^2_{i}) ,
\end{equation}
which depend on the Fermi momentum $k_{F}$. The coefficients $a_i$, 
$b_i$, $c_i$, and $\beta_i$ are given in Ref.~\cite{YNG}. The interaction 
matrix elements of the G-matrix derived from the chiral LO $YN$ potential 
\cite{YNLO} $G_{ijkl}(\Omega) =  \langle ij | \hat{G}(\Omega) | kl  
\rangle$ are obtained by solving a first approximation of the 
Bethe-Goldstone equation 
 \begin{eqnarray}
G_{ijkl}(\Omega) = && V^{N\Lambda-N\Lambda}_{ijkl}\label{Bethe}\\
&-& \sum_{mn} V^{N\Lambda-N\Sigma}_{ijmn} \frac{1}{(\epsilon^{N}_m +
\epsilon^{\Sigma}_n)-\Omega} V^{N\Lambda-N\Sigma}_{klmn}\,,\nonumber
\end{eqnarray}
where $\Omega$ is a parameter of the G-matrix, $V^{N\Lambda-N\Lambda}$ 
($V^{N\Lambda-N\Sigma}$) is the $N\Lambda-N\Lambda$ ($N\Lambda-N\Sigma$) 
channel of the LO YN interaction, and $\epsilon^{N}_m$ 
($\epsilon^{\Sigma}_m$) are nucleon ($\Sigma$) single-particle energies. 
The sum over single-particle states in Eq. (\ref{Bethe}) runs only over 
unoccupied nucleon states. The   single-particle energies 
$\epsilon^{\Sigma}_m$ were obtained by using the $N\Sigma-N\Sigma$ 
channel of the LO YN interaction in the equation analogous to that which 
we solve to obtain the single-particle levels of $\Lambda$ (see Eq. (4) 
in \cite{Ves1}). The parameter $\Omega$ in Eq. (\ref{Bethe}) is defined as 
\begin{equation}
\Omega = E_{\rm{av}} - (m_{\Sigma^0} - m_p)\,,  
\label{EAV}
\end{equation}
where $m_{\Sigma^0}$ ($m_p$) is mass of $\Sigma^0$ (proton),  
and $E_{\rm{av}}$ is either a free parameter or we can determine it as an 
average energy of occupied nucleons, i.e.,  
$E_{\rm{av}}=\sum_i \epsilon^N_i v_i$/A, where the coefficient $v_i = 1$ 
for the occupied single-particle states, and $v_i = 0$  otherwise. 
%
%
\section{Effects from kaon distortion and kinematics} 
Before we will show results for the heavy hypernuclei we discuss in more 
detail effects from re-scattering of kaons in the final state and from 
shifting the hypernucleus mass due to an excitation energy. 
The latter was used in our approach~\cite{archiv,fermi,structure} 
but has not been discussed yet. 

As shown in Eq.~(\ref{kaonwave}) the kaon distorted wave depends on 
the nucleon density and kaon-nucleon ($KN$) scattering amplitude. 
For the latter we utilize a separable form constructed in 
Ref.~\cite{mesons98}. This form was also used in the description of kaon 
scattering on nuclei~\cite{mesons95,CJP97}. 
Here and in previous DWIA calculations of 
electroproduction of hypernuclei~\cite{archiv,HYP2018,fermi,structure}   
we have utilized an improved version of the separable $KN$ amplitude 
which includes the partial waves up to $l=7$ and which can describe 
the available $KN$ data up to the invariant energy of about 2.4 GeV. 
This is especially important for the hypernucleus calculations at 
photon energies of about 2 GeV~\cite{archiv} as other $KN$ amplitudes, 
e.g., that in Ref~\cite{martin}, fail to describe the $KN$ scattering 
data at those energies. 
Note also that this new separable $KN$ amplitude is isospin dependent 
and therefore the distortion differs for various targets according 
to a number of protons and neutrons, e.g., in $^{40}$Ca and $^{48}$Ca. 

The nucleon density $\rho$ in Eq. (\ref{kaonwave}) is described by the 
Hartree-Fock (HF) form which is taken from the HF many-particle 
calculations consistently with the OBDME. However, in our previous 
calculations for the $p$-shell 
hypernuclei~\cite{archiv,HYP2018,fermi,structure} we used 
a phenomenological parametrization motivated by the harmonic 
oscilator model (HO)~\cite{DeVries,DThesis}  
\begin{equation}
 \rho_{\rm HO}(r)= [a_0+a_1(\alpha_N\,r)^2]\exp{[-(\alpha_N\,r)^2]}\,, 
 \label{roHO}
\end{equation}
where the parameters depend on the nucleus mass number A and 
the HO parameter $b_{\rm HO}$
\begin{eqnarray}
a_0&=&\frac{3A\,\alpha_N^3}{(A-1)\,\pi^{3/2}}\,,\ \ 
a_1=\frac{2(A-4)}{9}\,a_0\,,\nonumber\\
\alpha_N&=&\sqrt{\frac{A}{A-1}}\,\frac{1}{b_{HO}}\,.
\end{eqnarray}
In the following discussion we compare influences of the HF and HO 
forms of the nucleon density on the cross sections. 

Remind that in our approach the kaon distortion is included via 
the eikonal approximation with the kaon-nucleus first-order optical 
potential which can be regarded as a good approximation because 
the considered kaon momenta are large (1-2 GeV) and the depth 
of the kaon-nucleus optical potential is moderate ($\approx 50$ MeV) 
due to relatively weak kaon-nucleon interaction. The latter makes 
a long mean-free-path of K$^+$ in the nuclear medium ($\approx 5$ fm). 
Moreover, the kaon angles considered in our calculations are very small, 
typically a few degrees. 
One can, of course, consider improvements of the eikonal approximation, 
e.g. including the Coulomb interaction~\cite{Capel} or corrections 
needed at smaller kaon momenta~\cite{Buuck}, but these extensions are 
beyond the scope of the present work. Here we only estimate 
Coulomb effects in the scattering of kaons on nuclei. 

Contributions from the Coulomb interaction (CI) in the total 
cross section of the kaon-nucleus scattering are shown in Fig.~2 
of Ref.~\cite{CJP97}. Those calculations were performed in the momentum 
space using a relativistic equation with the first-order kaon-nucleus 
optical potential. In that figure one can observe that inclusion of CI 
suppresses the cross section and the effect diminishes with rising 
kaon momentum. At the kaon lab momentum of 0.5 GeV/c the effect 
amounts to 6, 10, and 12 percent for the nuclei $^{12}$C, 
$^{28}$Si, and $^{40}$Ca, respectively~\cite{CJP97}. 
At higher momenta, e.g. 1 GeV/c, the effect is only 3\% for $^{40}$Ca.  
Therefore, in the electroproduction with the kaon momentum of 
about 1.3 GeV/c (and still larger) one can estimate that the 
suppression of the kaon-nucleus cross section due to CI amounts 
to about 1\% for $^{40}$Ca and a few percent for heavier nuclei.  
In the following discussion the kaon distortion in the electroproduction 
cross sections is included only via the strong interaction as given 
in Eq.~(\ref{kaonwave}).

In the case of light p-shell hypernuclei like $^{12}_{~\Lambda}$B the 
electroproduction cross section is decreased due to the kaon final-state 
re-scattering by 45\% for the ground-state doublet with $\Lambda$ in the 
$s$ orbit and by 34\% for the multiplet at about 11 MeV with $\Lambda$ 
in the $p$ orbit. In this case, the distortion is calculated using the HO 
density with $b_{\rm HO}=1.621$ fm and the suppression of the cross 
sections for the multiplets depends only very weakly on the many-particle 
approach and the effective $YN$ interaction, particularly in the considered 
case of the shell model with the $p$-shell optimized 
interaction~\cite{JohnM,archiv} and the TD$_\Lambda$ approach with the 
Nijmengen F interacion~\cite{structure}. 

\begin{figure}[htb]
\begin{center}
\includegraphics[width=6.2cm,angle=270]{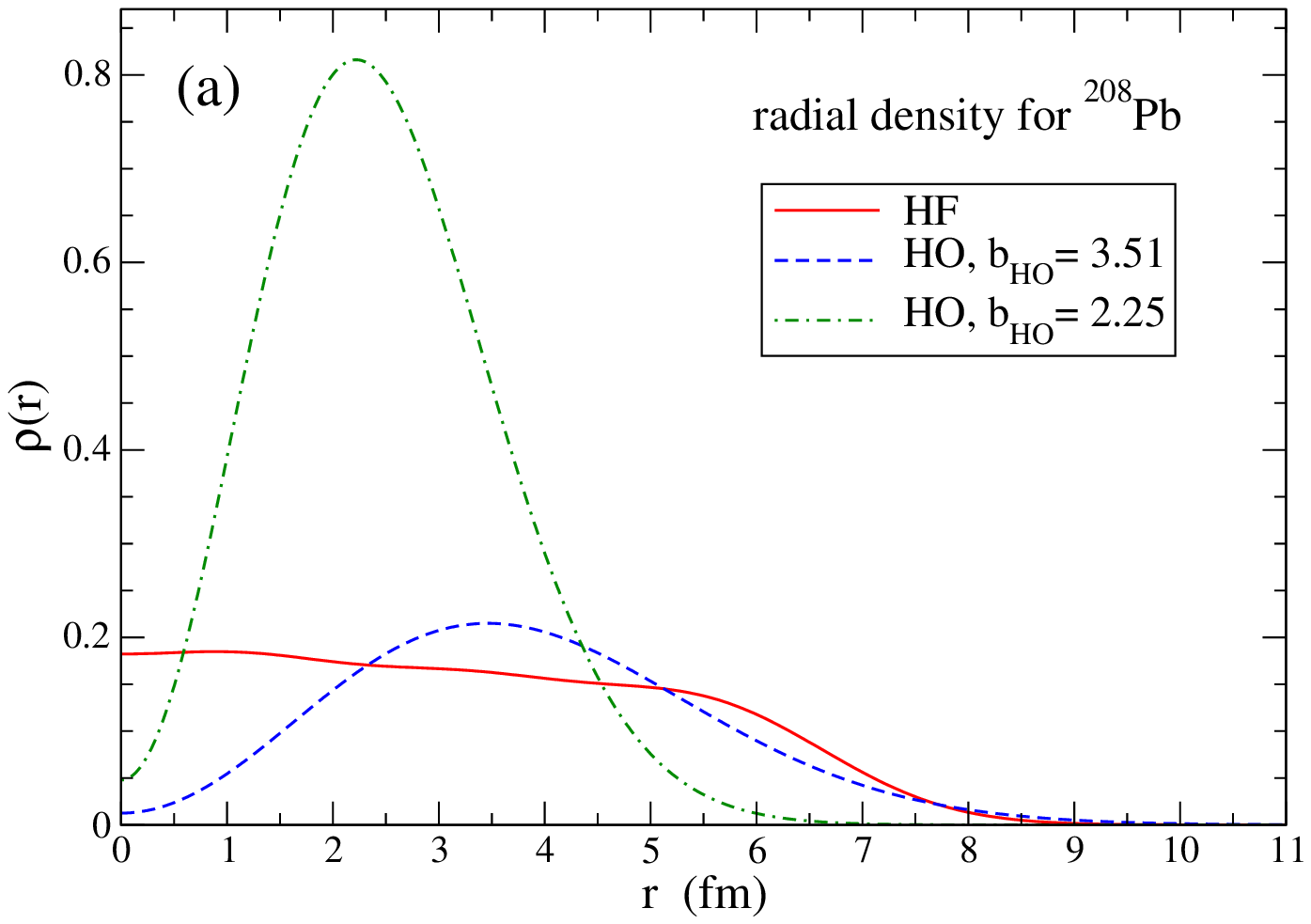}
\includegraphics[width=6.2cm,angle=270]{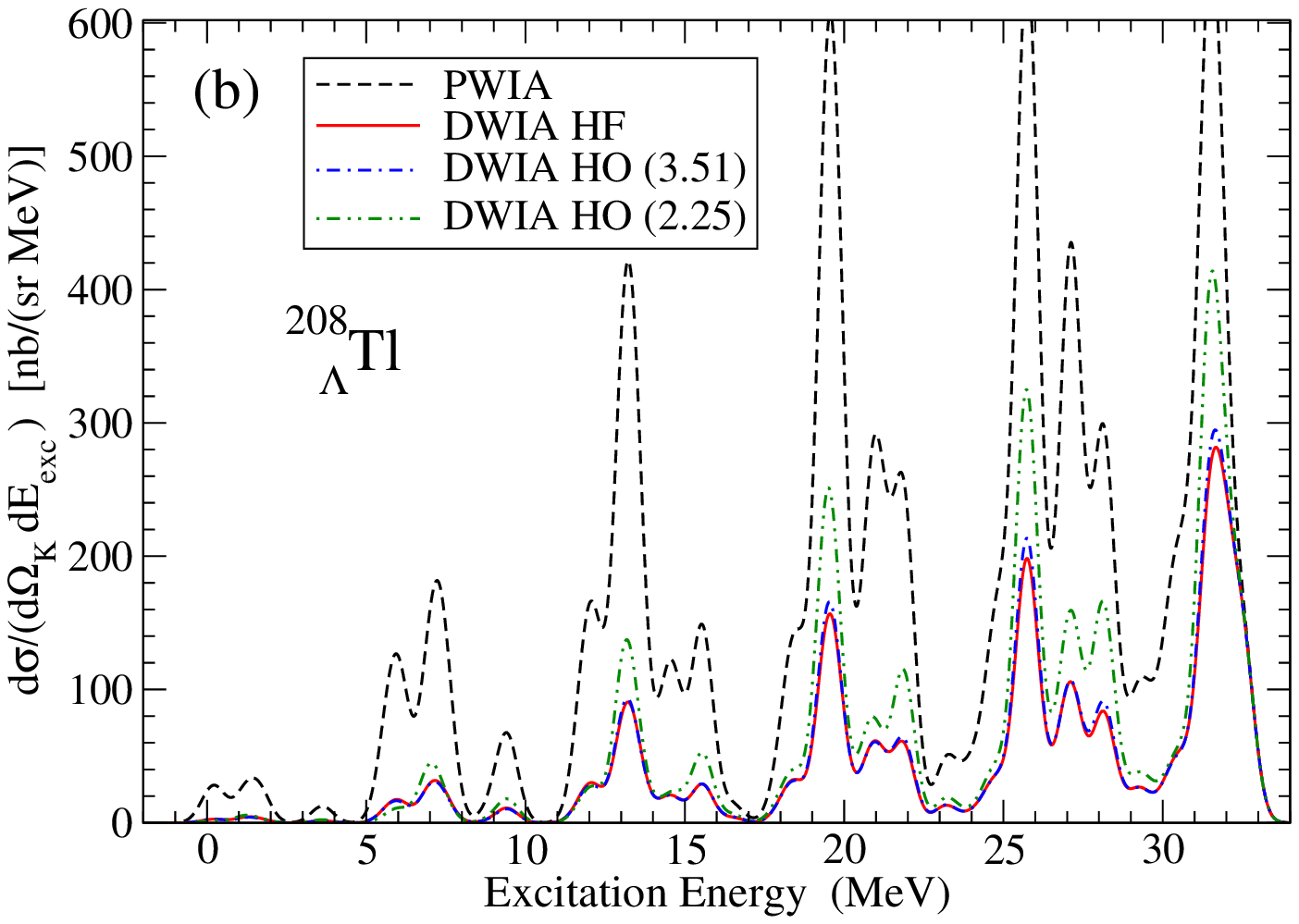}
\end{center}
\caption{Effects from kaon distortion for $^{208}_{\;\;\;\Lambda}$Tl.
The panel (a) shows radial nucleon densities from the harmonic-oscilator 
parametrization (HO) with $b_{\rm HO}=3.51$ and 2.25 fm and 
the Hartree-Fock calculation (HF), both normalized to A = 208. 
Panel (b) compares the excitation spectra calculated in PWIA and DWIA. 
In the latter the kaon-nucleus optical potential is calculated using 
the HO and HF nucleon densities shown in (a). The calculations are 
performed in the Tamm-Dancoff (TD$_\Lambda$) formalism using the 
Nijmegen F YNG interaction with the Fermi momentum $k_{F} =$ 1.34 
fm$^{-1}$ and the BS3 amplitude in the optimum on-shell approximation. 
The curves are plotted with FWHM = 800 keV.}
\label{208Pb-distorse}
\end{figure}

In Figure~\ref{208Pb-distorse} we show effects of kaon distortion in 
the case of heavy hypernucleus $^{208}_{\;\;\;\Lambda}$Tl calculated 
with three different nucleon densities. 
The aim is to demonstrate dependence of the distortion effects 
on behavior of the optical potential given by the density. 
The Hartree-Fock (HF) and two harmonic-oscilator (HO) forms of 
the densities are shown in part (a), the latter with $b_{\rm HO}=$ 3.51 
fm and 2.25 fm. It is clear that the HO densities do not provide 
a realistic description, mainly in the central region of the nucleus. 
However, as one can observe in part (b) the suppression of the 
excitation spectrum caused by the HF and HO(3.51) densities do not 
differ too much, suggesting that the effect is given mainly by 
a behavior of the kaon-nucleus optical potential in the peripheral 
region whereas its behavior inside the nucleus is irrelevant. 

A characteristic related to the nucleon density is the mean 
nucleus radius. The nucleus radius calculated with the HF density 
is $R_{\rm HF}$ = 5.26 fm and those with the HO densities are 5.22  
and 3.36 fm for $b_{\rm HO}$ = 3.51 and 2.25 fm, respectively. 
The value of $R_{\rm HF}$ is in a good agreement with the values 
extracted from data on electron scattering, $R_{\rm exp}=5.50$ 
fm~\cite{DeVries} suggesting that the HF form of the density 
is realistic. Note also that in Fig.~\ref{208Pb-distorse} we 
can see that the HO(3.51) density is quite close to the HF one 
in the peripheral region and the corresponding distortions are 
almost the same. 
A more detailed study of distortion effects in electroproduction of 
$p$- and $sd$-shell hypernuclei can be found in Ref.~\cite{DThesis}.  

We can conclude that the distortion effects amount to 40--60\% for 
$^{40}_{~\Lambda}$K and 50--80\% for $^{208}_{~~\Lambda}$Tl and that 
they depend on behavior of the nucleus density in the peripheral region 
rather than in the nuclear interior. Another observation is that the 
hypernucleus states with a more deeply bound $\Lambda$ hyperon, e.g.,  
in the $s$ orbit, are more strongly affected by the kaon distortion 
than the states with $\Lambda$ in higher lying orbits, e.g. $d$.  
Note also that the different strength of the kaon distortion for various 
states changes the shape of multiplets in the spectrum as observed 
in Fig.~\ref{208Pb-distorse}.  
In our present calculations we use the HF densities which are consistent 
with the experimental observables providing a realistic optical 
potential and which are obtained in the many-particle calculations 
consistently with the values of OBDMEs.
%
%

As we mentioned above, in determination of kinematics we use the 
hypernucleus mass shifted by the excitation energy of given state: 
$M^*_H= M_H^0 + E^*$ where $M_H^0$ and $E^*$ are the hypernucleus 
ground-state mass and excitation energy, respectively. 
Even if this mass shift is relatively very small it can change values 
of some particle momenta even by a few percent and therefore the 
elementary amplitude and the radial integrals can notably change their 
values. 
In Table~\ref{shift} we show how a relatively tiny shift of the mass 
propagates into variations of the momenta, the elementary amplitude, and the 
radial integral, and finally how it changes the cross section. The results, 
calculated in PWIA at photon energy $E_\gamma=1.5$ GeV, are shown for 
two states of $^{208}_{~~\Lambda}$Tl with the same spin to see variations 
of the effect in dependence on $E^*$. 
Whereas changes of the momenta are proportional to 
the excitation energy, the radial integral reveals different effects and 
therefore the cross section can be less suppressed for the larger shift. 
The value of the radial integral is given by the overlap of the proton and 
$\Lambda$ single-particle wave functions and the function proportional 
to the spherical Bessel function in the PWIA,  
$F_{LM}(\Delta B\xi)\sim j_L(\Delta B\xi)$, see Eq.~(\ref{radial}). 
A few percent change of $\Delta$ can vary the overlap of the functions  
resulting in a quite different value of 
${\cal R}^{LM}_{\alpha^\prime\alpha}$. It is also remarkable in the table 
that the particle momenta decrease due to the mass shift but the momentum 
transfer $\Delta$ rises and that the proton momentum $|\vec{p}_{opt}|$ 
significantly changes which then markedly affects the value of the 
elementary amplitude.  
The changes are also present in the transversal part (d$\sigma_T$) of the 
full cross section (d$\sigma$). Note that dependence of the cross 
sections on the momentum transfer $\Delta$ was also studied in the DWIA 
calculations of photoproduction of  $^{208}_{\;\;\;\Lambda}$Tl in 
Ref.~\cite{MotobaJPS} showing noticeable effects. 
%
%
\begin{table}[hbt]
\begin{center}
\begin{tabular}{ccc}
\hline
 & \multicolumn{2}{c}{states}\\
 \hline
                  & (11.964, $8^-$) & (18.201, $8^+$) \\
                  \hline
 $M_H^*$            &  0.006      &  0.009        \\
$|\vec{P}_K|$     &   -1.0      &   -1.6        \\
$\Delta$  &    3.0      &    4.6        \\
$|\vec{p}_{opt}|$ &  -15.7      &  -23.5        \\
Re${\cal F}^1_{00}$&  -1.2      &   -4.5        \\
Im${\cal F}^1_{00}$&  -31.4     &  -44.4        \\
${\cal R}^{LM}_{\alpha^\prime\alpha}$&${\cal R}^{70}_{\alpha^\prime\alpha}: \,-3.7$ & ${\cal R}^{80}_{\alpha^\prime\alpha}:\,20.2$ \vspace{1mm}\\
d$\sigma$  &    -16.0   &   -10.6 \\
d$\sigma_T$&    -13.2   &   -9.6 \\
\hline
\end{tabular}
\caption{Changes, in \% with respect to the ground-state values,  
are due to the mass shift for two states of $^{208}_{~~\Lambda}$Tl: 
($E^*$[MeV], $J_{\rm H}^P$). The changes are shown for the kaon 
momentum $|\vec{P}_K|$, momentum transfer $\Delta=|\vec{\Delta}|$, 
proton optimum momentum $|\vec{p}_{opt}|$, the spin-flip elementary 
amplitude with zero projections and the relevant radial integrals 
for dominant transitions $0h_{11/2}\to 0d_{5/2}$ and 
$0h_{11/2}\to 0f_{5/2}$. The full (d$\sigma$) and transversal 
(d$\sigma_T$) cross sections~\cite{fermi} decrease by 10--16\%. }
\label{shift}
\end{center}
\end{table} 

\begin{figure}[htb]
\begin{center}
\includegraphics[width=6.2cm,angle=270]{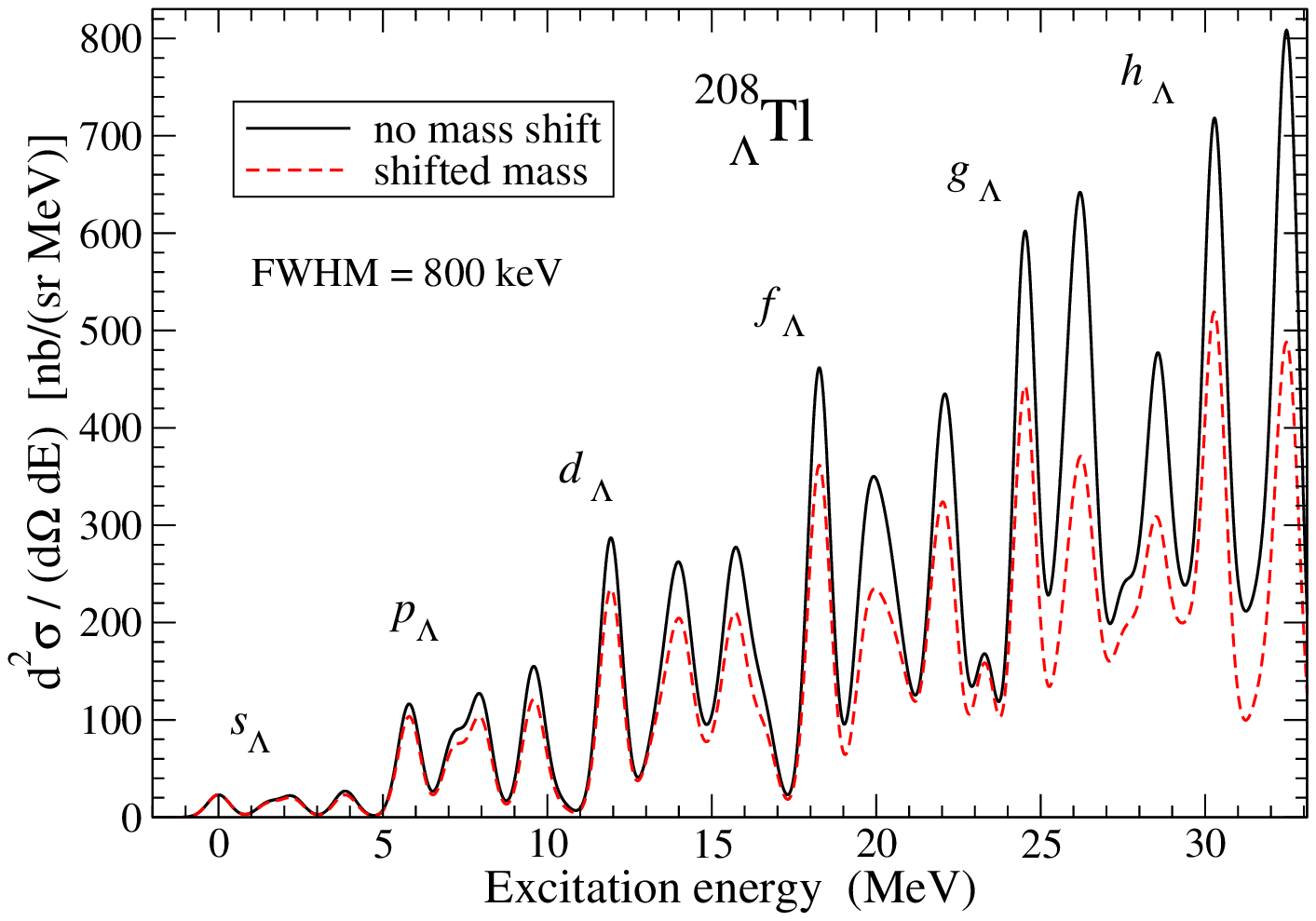}
\end{center}
\caption{The excitation spectra of $^{208}_{\;\;\;\Lambda}$Tl calculated 
in PWIA without and with the shift of hypernucleus mass due to the 
excitation energy. The calculations are for $E_\gamma=1.5$ GeV and 
$\theta_{K\gamma}=8^\circ$ with the BS3 elementary amplitude in the 
optimum on-shell approximation. }
\label{208Pb-shift}
\end{figure}

Despite the fact that the cross sections for various states can change  
in different way, as one can see in Table~\ref{shift}, the overall effect  
in the excitation spectrum is monotonically rising with the excitation 
energy as shown in Fig.~\ref{208Pb-shift}. Therefore, both the
$d_\Lambda$ and $f_\Lambda$ peaks change similarly by about 16\% 
even if the cross section for the state (18.201, $8^+$) changes only 
by 10.6\%. This is because of dominance of the first state (11.964, $8^-$) 
in the $d_\Lambda$ peak whereas the $f_\Lambda$ peak is dominated by the 
(18.316. $9^+$) state with the cross section changed by 16.3\%. 
%
%

We can conclude that the effects from the hypernucleus-mass shift are quite 
important correction to the cross section of high-lying hypernucleus 
states. In the considered kinematical region the mass shift suppresses 
the cross sections. 

%
%
\subsection{On-shell approximation}
In the impulse approximation the hypernucleus production amplitude is 
given by an integral over the momentum of the target proton from the 
elementary production amplitude and the transition many-particle matrix 
element, see Eq.(5) in Ref.~\cite{fermi}. This Fermi-momentum averaging 
integral is factorized assuming the optimal factorization 
approximation~\cite{fermi}, where the elementary amplitude is evaluated 
at an effective proton momentum. This effective momentum is to be chosen 
in a model calculation.

In previous works the effective momentum was mostly chosen to be zero, 
in the so called frozen-proton 
approximation~\cite{archiv,Motoba2010,NPA881,MotobaJPS,HYP2018}. 
Effects of using other values of the effective momentum (the Fermi 
motion effects) were studied by Mart {\it et al.} in electromagnetic 
production of the hypertriton~\cite{Mart1,Mart2}. We will show 
the Fermi motion effects in the case of a heavy hypernucleus in the 
next subsection. 

\begin{figure}[htb]
\begin{center}
\includegraphics[width=2.4cm,angle=270]{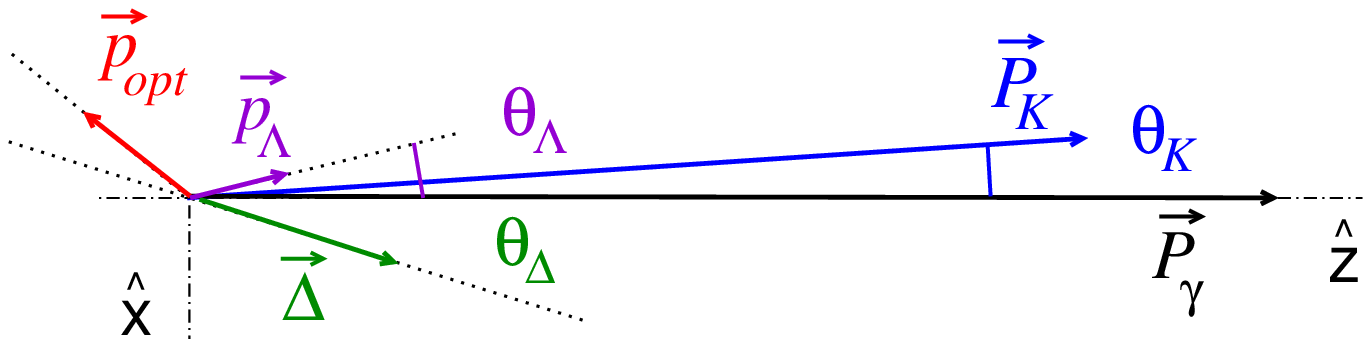}
\end{center}
\caption{Kinematics of the elementary process in the coplanar case 
with the axis $\hat{\sf y}$ oriented rearwards the page. The angle 
$\theta_{\Delta p}$ between the proton optimum momentum $\vec{p}_{opt}$ 
and the momentum transfer $\vec{\Delta}=\vec{P}_\gamma -\vec{P}_{\rm K}$ 
has a general value.}
\label{coplanar}
\end{figure}

In our recent analysis~\cite{fermi} we suggested using the optimum 
on-shell approximation with an optimum proton momentum 
($\vec{p}_{opt}$) as the effective value. This optimum momentum 
is calculated from the two-body energy conservation in 
the laboratory frame (the nucleus is in rest) 
\begin{eqnarray}
E_\gamma +\sqrt{m_p^2+(\vec{p}_{opt})^2} &=&
\sqrt{m_K^2+|\vec{\sf P}_{\sf K}|_{mb}^2}\ +\nonumber\\ 
&+&\sqrt{m_\Lambda^2+(\vec{\Delta} + \vec{p}_{opt})^2}\,,
\label{2bec}
\end{eqnarray}
where $m_p$, $m_K$, and $m_\Lambda$ are the proton, kaon, and 
$\Lambda$ masses, respectively, $E_\gamma$ is the photon energy, 
and $|\vec{P}_K|_{mb}$ is the magnitude of the kaon lab momentum 
computed from the many-body energy conservation. This equation assures 
that the elementary amplitude is on shell but one still has to chose 
a value of the angle $\theta_{\Delta p}$ between the proton momentum 
and the momentum transfer $\Delta$, which introduces a residual 
indeterminacy in this approach. 
In previous calculations~\cite{fermi,structure}, we had chosen 
$\theta_{\Delta p}= 180^\circ$ because then both proton and $\Lambda$ 
momentum received the minimum value at given kinematics. 
Kinematics in the coplanar case 
($\Phi_K=\Phi_p$) is depicted in Fig.~\ref{coplanar}. 
%
%

In the present work we elaborate the optimum on-shell approximation 
assuming that the magnitude of $\vec{p}_{opt}$ equals the mean momentum 
of the proton in the single-particle state of the dominant transition. 
This mean momentum, $p_{mean}=\sqrt{2\mu\,\langle\,T_{kin}\,\rangle}$ 
with the reduced mass $\mu$, is calculated for each hypernucleus 
state from the proton single-particle wave function that corresponds 
to the dominant value of the OBDME. The mean value of the kinetic-energy 
operator  $\langle\;T_{kin}\,\rangle$ is evaluated directly from its 
definition differentiating numerically the radial wave functions. 

Having fixed the magnitude of the proton momentum, 
$|\vec{p}_{opt}|=p_{mean}$, we can calculate the angle 
$\theta_{\Delta p}$ from Eq.~(\ref{2bec}) and if it acquires a physical 
value, i.e., $\cos\theta_{\Delta p} \geq -1.0$, we can use this fully 
determined value of the proton momentum 
$\vec{p}_{opt}=(p_{mean},\theta_{\Delta p},\Phi_K)$ in 
the hypernucleus calculations. 
In the case the angle is nonphysical, $\cos\theta_{\Delta p} < -1.0$, 
we can calculate $|\vec{p}_{opt}|$ from Eq.~(\ref{2bec}) with 
$\cos\theta_{\Delta p}=-1.0$ as before. This provides the minimal value 
of $|\vec{p}_{opt}|$ at given kinematics, and because $p_{mean}$ is 
still smaller, this value of $|\vec{p}_{opt}|$ is the best substitution 
for $p_{mean}$. 

We summarize the variants of  the optimum on-shell approximation:\\ 
\hspace*{3mm} 
(a) {\it the previous version in Refs.~\cite{fermi,structure}:} 
the angle is chosen as $\theta_{\Delta p}=180^\circ$ and the magnitude  
of $\vec{p}_{opt}$ calculated from Eq.~(\ref{2bec});\\
\hspace*{3mm} (b) {\it the present elaborated version:} the magnitude 
is specified as $|\vec{p}_{opt}|=p_{mean}$ and the angle 
$\theta_{\Delta p}$ calculated from Eq.~(\ref{2bec}). 
If the angle is physical we use this variant otherwise we use 
the variant a).\\
In both variants the proton azimuthal angle is taken to equal that of 
the kaon, $\Phi_p=\Phi_K$. 

In the following we are going to use the variant (b) and denote it as  
``on-shell approximation''. 
Note that in this variant the proton effective momentum is fully 
determined. 
Since many hypernucleus states, especially the low lying states, 
have nonphysical values of $\theta_{\Delta p}$ and the values of 
$p_{mean}$ are in general close to $|\vec{p}_{opt}|$ with 
$\cos\theta_{\Delta p}=-1.0$, the new and old variants of the optimum 
on-shell approximation do not differ too much.
 
Recall also that the on-shell approximation is based on the optimal 
factorization, which replaces the full Fermi averaging integral. 
However, in this approximation one can utilize the on-shell 
elementary amplitude determined in analysis of the elementary 
production process.
%
%
\subsection{Fermi motion effects}
In 2008 Mart and Ventel found that the Fermi motion effects are 
essential for a correct description of the hypertriton 
electroproduction~\cite{Mart2}.
Dependence on various values of the proton effective momentum of 
the cross sections in electroproduction of p-shell hypernuclei  
was discussed in Ref.~\cite{fermi}.  
In Figures~\ref{208Pb-adep} and \ref{208Pb-edep} we show these 
Fermi motion effects for several states ($E^*$ [MeV], $J_{\rm H}^P$) 
of the heavy hypernucleus $^{208}_{~~\Lambda}$Tl in dependence on 
the kaon angle with respect to the electron beam $\theta_{\rm Ke}$ 
and on the photon laboratory energy $E_\gamma$, respectively. 
%
%
\begin{widetext}
\begin{center}
\begin{figure}[bth]
\includegraphics[width=11.2cm,angle=270]{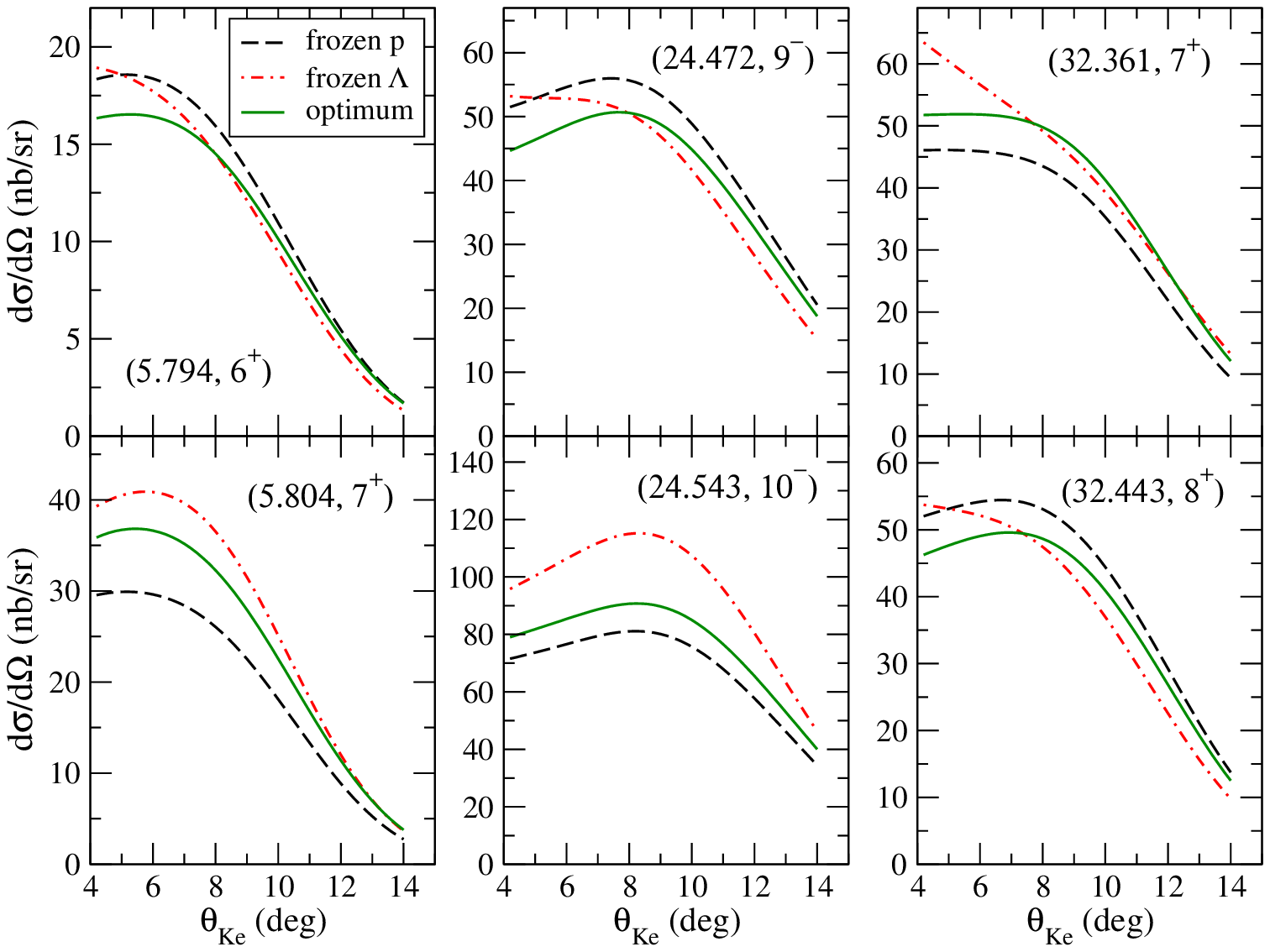}
\caption{Angular dependence of the cross sections for selected states 
($E^*$ [MeV], $J_{\rm H}^P$) in electroproduction of 
$^{208}_{\;\;\;\Lambda}$Tl. The results are shown for 
various values of the effective proton momentum, $\vec{p}_{eff}=0$ 
(frozen p), $\vec{p}_{eff}=-\vec{\Delta}$ (frozen $\Lambda$), and  
$\vec{p}_{eff}=\vec{p}_{opt}$ (optimum). The calculations were performed 
with the elementary amplitude BS3~\cite{SB} at $E_\gamma= 1.5$ GeV.  
}
\label{208Pb-adep}
\end{figure}
\end{center}
\end{widetext}

The biggest differences of the cross sections, mostly observed between the 
boundary cases of ``frozen'' proton and $\Lambda$, are at small kaon angles 
amounting about 20--30\%. The energy dependence of the effects reveals a 
resonant-like structure in some cases, where more pronounced effects are 
observed for energies below 2 GeV.
Similarly as in the case of lighter hypernuclei one can see that 
the nature of the effects is very similar for the states 6$^+$, 
9$^-$, and 8$^+$ and for 7$^+$ and 10$^-$, particularly the sequence 
of the curves is almost identical. The similarity of the effects 
inside these two groups of states is given by the selection rule 
discussed in Ref.~\cite{fermi}. 
This rule says that contributions from the longitudinal mode of the 
virtual photon ($\lambda=0$) are important only for the states with 
specific spin and parity. 
Let us show this in the example: 
both states (5.794, 6$^+$) and (5.804, 7$^+$) are dominated by the 
proton$\to\Lambda$ transition $0h\to 0p$ which means that $L=6$ in 
Eq. (\ref{amplitude-3}). As the leading contribution to the reduced 
amplitude with $\lambda=0$ comes from the spin-flip elementary amplitude 
with $S=1$ and $\eta=0$ and the biggest value of the radial integral is 
for $M=0$ the corresponding Clebsch-Gordan coefficients in 
Eq.~(\ref{amplitude-3}) are ${\sf C}^{60}_{6010}=0$ and 
${\sf C}^{70}_{6010}=\sqrt{7/13}$. The longitudinal contributions in 
the case of $7^+$ then make the frosen-$\Lambda$ cross section mostly 
larger than that for the case with the stationary proton contrary to the 
case of the state $6^+$. This longitudinal contribution also makes the 
pronounced resonant structures in the energy dependent cross sections 
in Fig.~\ref{208Pb-edep}. These observations demonstrate the importance 
of the contributions from the longitudinal and transverse-longitudinal 
interference terms in the electroproduction cross section~\cite{fermi}. 
%
%
\begin{widetext}
\begin{center}
\begin{figure}[htb]
\includegraphics[width=11.2cm,angle=270]{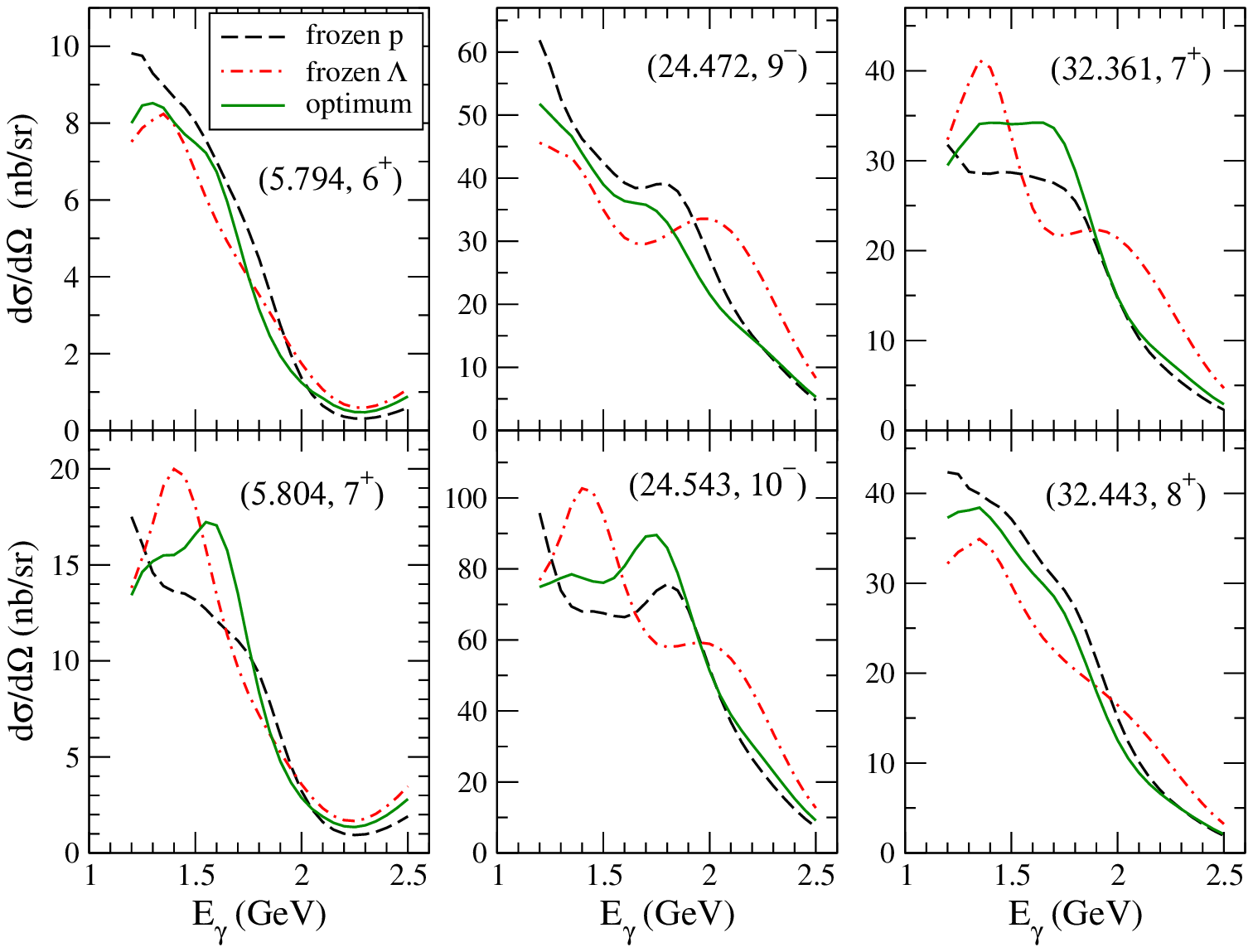}
\caption{The same as in Fig.~\ref{208Pb-adep} but for dependence 
on the photon lab energy at $\theta_{\rm Ke}=11^\circ$.}
\label{208Pb-edep}
\end{figure}
\end{center}
\end{widetext}

The results with the optimum momentum mostly lie within the limit cases 
with the frozen proton and $\Lambda$ which is given by values of the 
proton and $\Lambda$ momenta. For example, at $E_\gamma =1.5$ GeV and 
$\theta_{Ke}=11^\circ$ the proton and $\Lambda$ momenta for the state 
(5.804, 7$^+$) are, respectively, 0 and 392 MeV/c for the frozen proton, 
292 and 0 MeV/c for the frozen $\Lambda$, and 207 and 162 MeV/c for 
the optimum proton momentum case. Note that the results in 
Figs.~\ref{208Pb-adep} and \ref{208Pb-edep} are calculated with the 
elementary amplitude BS3~\cite{SB} which has a more abundant energy 
structure than the Saclay-Lyon amplitude as discussed in Ref.~\cite{fermi}. 
%
%
\section{Discussion of excitation spectra for medium- and  
heavy-mass hypernuclei}
\subsection{$^{40}_{~\Lambda}$K and $^{48}_{~\Lambda}$K}
The excitation spectra in electroproduction of $^{40}_{~\Lambda}$K and 
$^{48}_{~\Lambda}$K were already discussed in Ref.~\cite{structure}. 
It was shown how much the spectra depend on the Fermi momentum $k_F$, 
used to parametrize the YNG Nijmegen F interaction, and on the elementary 
amplitudes SLA and BS3. We have also compared the results obtained using  
two different approaches, namely TD$_\Lambda$ and the Equation of 
Motion Phonon Method (EMPM$_\Lambda$).  
These results were obtained using the chiral NNLO$_{sat}$ $NN+NNN$ 
potential~\cite{NN-NNN} and the kaon distortion was computed with 
the HO nucleus density and $b_{\rm HO}= 1.939$ fm~\cite{structure}. 

\begin{figure}[htb]
\begin{center}
\includegraphics[width=6.2cm,angle=270]{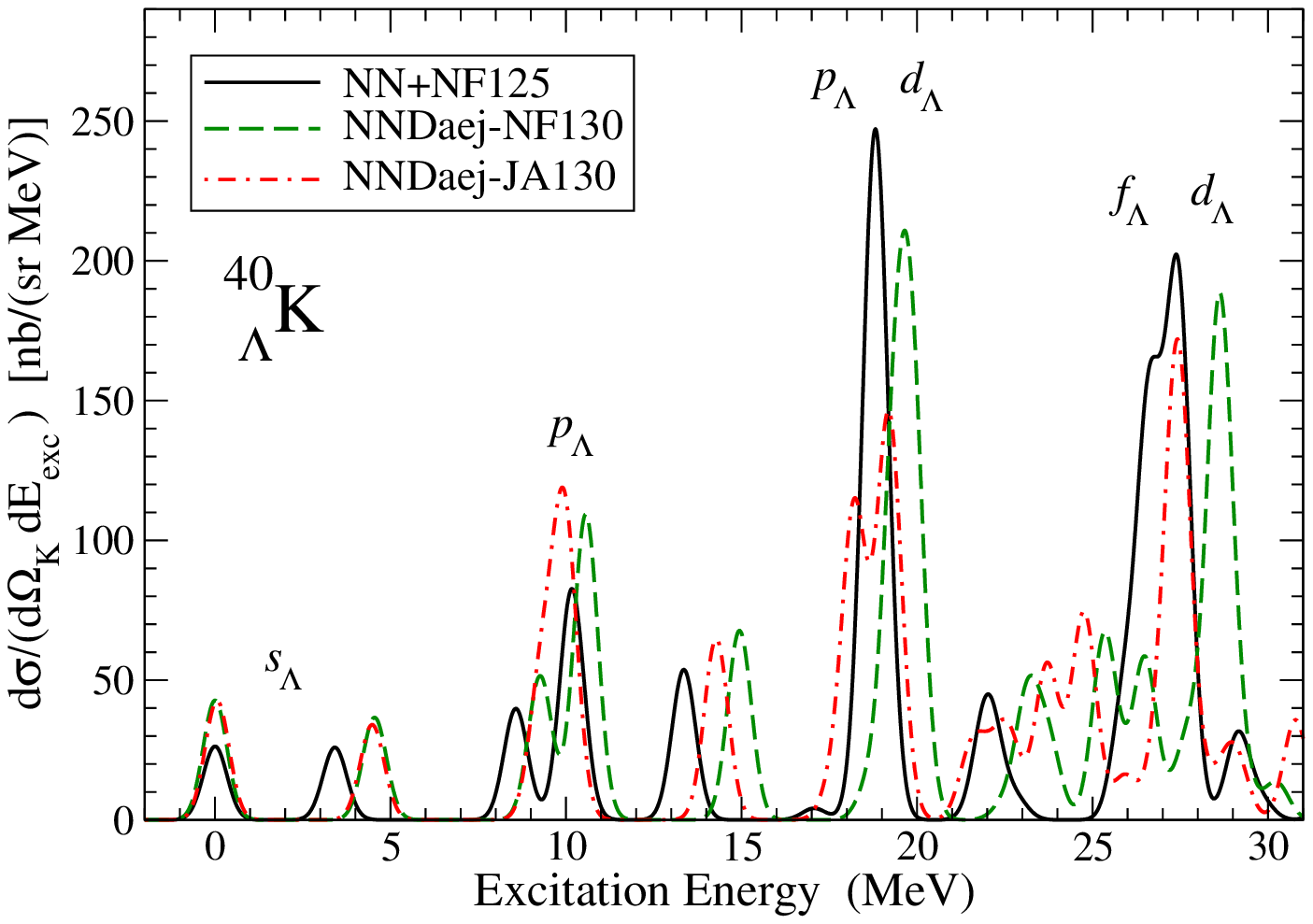}
\end{center}
\caption{The previous result for $^{40}_{~\Lambda}$K with the BS3 
amplitude in the optimum on-shell approximation~\cite{structure} 
(NN+NF125) is compared with the upgraded calculations using the 
D16+DDT interaction and the NF (NNDaej-NF130) and JA (NNDaej-JA130) 
YNG interactions, both with $k_F= 1.30$ fm$^{-1}$.
The kaon distortion is included using the HF density of $^{40}$Ca. 
The curves are plotted with FWHM = 800 keV.}
\label{40Ca-NN-YNG}
\end{figure}

\begin{figure}[htb]
\begin{center}
\includegraphics[width=6.2cm,angle=270]{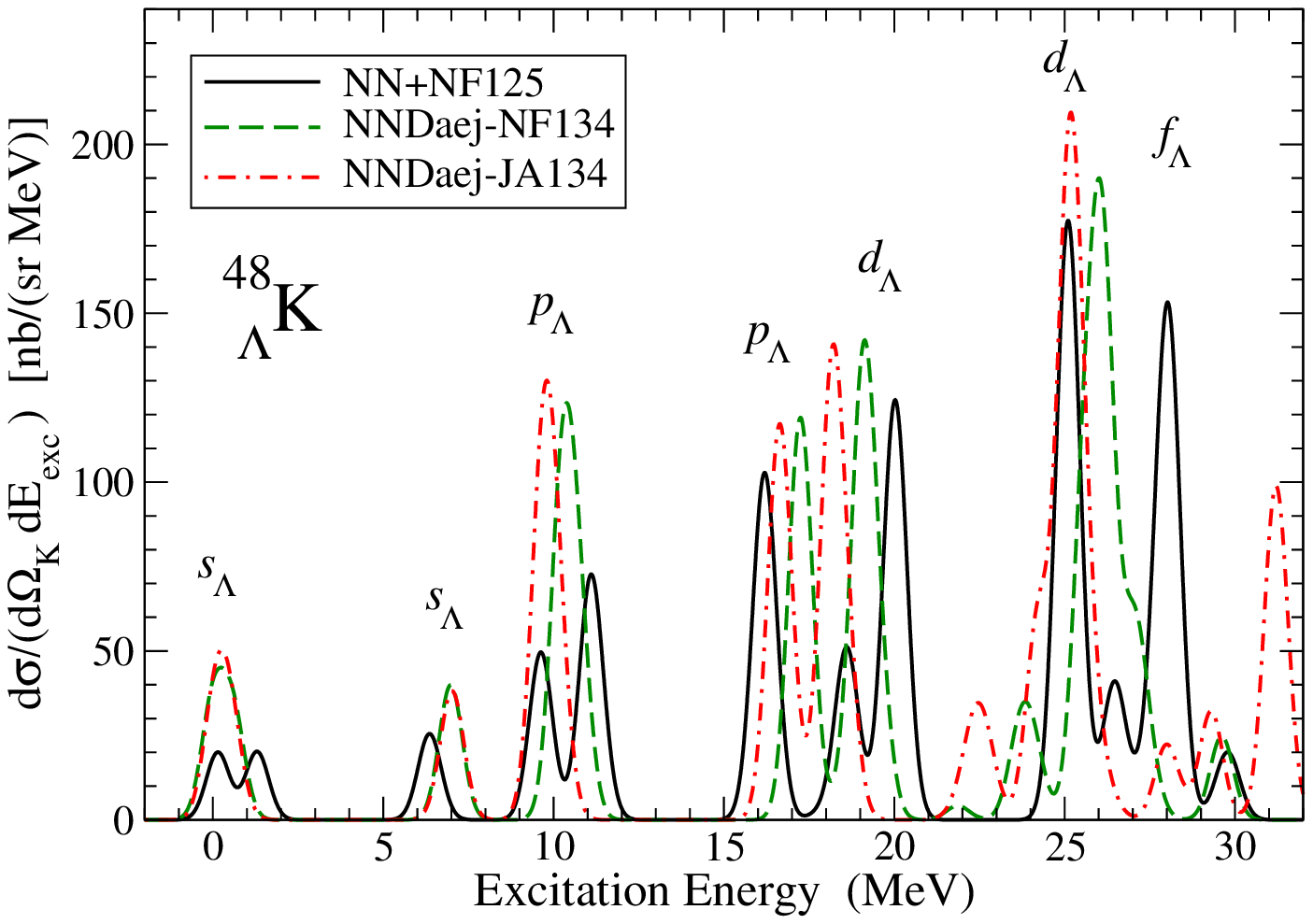}
\end{center}
\caption{The same as in Fig.~\ref{40Ca-NN-YNG} but for 
$^{48}_{~\Lambda}$K and $k_F= 1.34$ fm$^{-1}$.}
\label{48Ca-NN-YNG}
\end{figure}

In the present paper we show results calculated in the TD$_\Lambda$ 
approach using the D16+DDT interaction with 
$C_{\rho} = 2000$ MeV$ \cdot$fm$^6$ in Eq.~(\ref{DD}) and the kaon 
distortion is included using the optical potential constructed with 
the HF densities of the target nuclei obtained in the HF approach. 
The upgraded results calculated with the BS3 amplitude in the on-shell 
approximation and with two YNG interactions, J\"ulich A (JA) and 
Nijmegen F (NF), are compared with the previous calculations 
in Figs.~\ref{40Ca-NN-YNG} and \ref{48Ca-NN-YNG}. 
Note that the old results were calculated with $k_F=$ 1.25 fm$^{-1}$ 
and the new ones with 1.30 and 1.34 fm$^{-1}$ for $^{40}_{~\Lambda}$K 
and $^{48}_{~\Lambda}$K, respectively. Differences in the $NN$ 
interaction, the value of $k_F$, and using the JA YNG interaction 
result in changes in the excitation spectra, where the production strength 
is re-distributed into diverse structures. However, the magnitudes 
of the peaks, which are sensitive especially to the elementary 
amplitude, kinematics, and the kaon distortion, mostly do not differ 
too much for the NF and JA interactions with the given $k_F$.

The results presented in Figs.~\ref{40Ca-NN-YNG} and \ref{48Ca-NN-YNG} 
demonstrate a variety of predictions for the experiment E12-15-008 
in preparation at Jefferson Lab. Even if the natures of the NF and JA 
interactions differ the predicted excitation spectra reveal only mild 
differences, at least in the region of the $s_\Lambda$ -- 
$d_\Lambda$ peaks. Therefore only very good quality data (small FWHM) 
will be able to distinguish these two YNG interactions. 
%
%
%
%

Changing the NN interactioon and using larger Fermi momenta $k_F$ 
also results in different values of the $\Lambda$ s-wave binding 
energies ($B_\Lambda$) with respect to the results presented in 
Ref.~\cite{structure}. The binding energy calculated with the 
Nijmegen F YNG interaction is a bit larger for $^{40}_{~\Lambda}$K: 
18.62~\cite{structure} $\to$ 18.95 MeV (this work) but smaller for 
$^{48}_{~\Lambda}$K: 20.01~\cite{structure} $\to$ 19.72 MeV (this work). 
The new values accord quite well with the binding energies obtained 
by Friedman and Gal~\cite{FG}, 18.70 and 19.78 MeV for 
$^{40}_{~\Lambda}$K and $^{48}_{~\Lambda}$K, respectively. 
Note, however, that the $s_\Lambda$ -- $p_\Lambda$ excitation energy 
comes out still larger by about 1.5--2 MeV in our approach than in 
the calculations with the density dependent optical potential 
in Ref.~\cite{FG}. 

The $B_\Lambda$ calculated with the J\"ulich A interaction is 16.83 
and 17.83 MeV for $^{40}_{~\Lambda}$K and $^{48}_{~\Lambda}$K, 
respectively, which are smaller values that those for the Nijmegen F 
interaction suggesting that JA is weaker than NF. 
%
%
\subsection{$^{52}_{~\Lambda}$V}
We also provide predictions of the excitation spectrum in the reaction 
$^{52}$Cr($e,e^\prime K^+$)$^{52}_{~\Lambda}$V and compare them with 
the older calculation in Ref.~\cite{NPA881}. This older result presented 
in Ref.~\cite{NPA881}, Fig.~7, is for photoproduction in kinematics 
$E_\gamma = 1.3$ GeV and $\theta_{K\gamma} = 3^\circ$ and with the SLA 
amplitude in the frozen-proton approximation.  
The spectrum clearly shows separated multiplets with the $\Lambda$ 
hyperon in the $s$, $p$, $d$, and $f$ orbit. The major peaks are based 
on the conversion of protons in the $0f_{7/2}$ orbit into the $\Lambda$ 
bound in the given $l_\Lambda$ orbits. Note that the dominant 
unnatural-parity peaks ($4^-$, $5^+$, $6^-$, $7^+$) are with 
$J_{max}=L_{max}+1$. The other well separated and important peaks are 
based on the proton hole $d_{3/2}^{-1}$~\cite{NPA881}. 
%
%
\begin{figure}[htb]
\begin{center}
\includegraphics[width=6.2cm,angle=270]{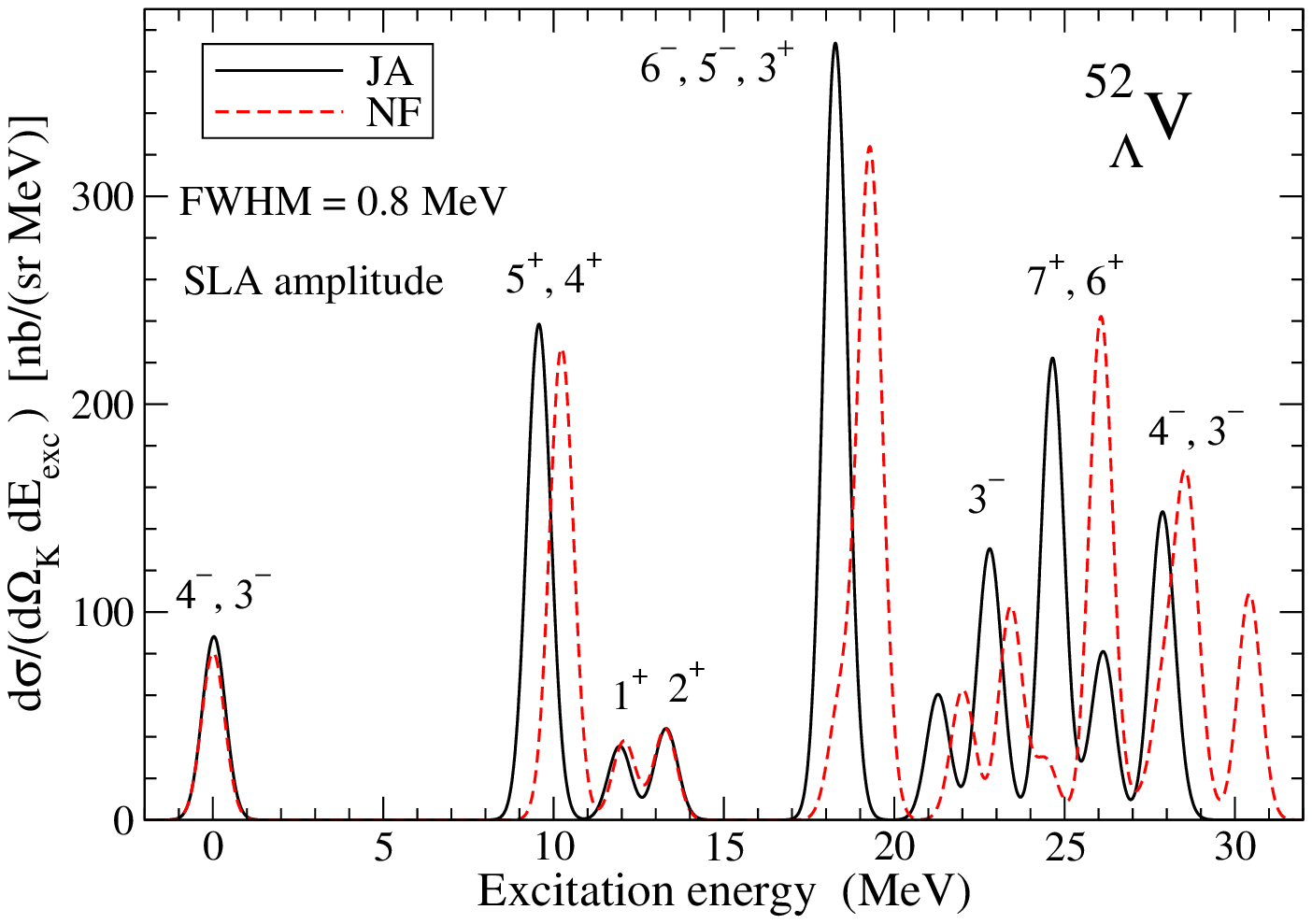}
\end{center}
\caption{Excitation spectra for electroproduction of $^{52}_{~\Lambda}$V 
calculated in the TD$_\Lambda$ approach with the J\"ulich A (JA) and 
Nijmegen F (NF) YNG interactions 
and with the elementary amplitude SLA in the frozen-proton approximation.}
\label{52Cr-YNG-SLA}
\end{figure}

The results presented here are for electroproduction in kinematics 
$E_i= 1.6$ GeV, $E_f= 0.3$ GeV, $\theta_e= 4.5^\circ$, 
$\theta_{Ke} =4.04^\circ$, and $\Phi= 180^\circ$, giving photoproduction 
kinematics $E_\gamma = 1.3$ GeV, $Q^2= 0.003$ (GeV/c)$^2$, 
$\varepsilon= 0.362$, $\theta_{\gamma e}= 1.04^\circ$, and 
$\theta_{K\gamma}= 3.0^\circ$, quite close to that in Ref.~\cite{NPA881}. 
The kaon distortion is included using the HF density for $^{52}$Cr and  
the OBDMEs are determined within the HF+TD$_\Lambda$ approach using 
the $NN$ interaction  D16+DDT with $C_{\rho} = 2000$ MeV$\cdot$fm$^6$ 
in Eq.(\ref{DD}) and JA and NF YNG interactions with $k_F=1.34$ fm$^{-1}$. 

In Figure~\ref{52Cr-YNG-SLA} we show results calculated with the SLA 
amplitude in the frozen-proton approximation ($p_{eff}=0$) and the JA and 
NF YNG interactions, which can be compared with the results in Fig. 7 of 
Ref.~\cite{NPA881}. One can observe that the main three peaks in the 
JA result are of similar magnitude and position as in Ref.~\cite{NPA881}. 
In the new calculations, these main peaks are especially formed by the 
following states ($E^*$ [MeV], $J_H^P$) dominantly populated by the 
single-particle transitions $nl_j\to n^\prime l^\prime_{j^\prime}$: 
(0.0 MeV, $4^-$) with $0f_{7/2}\to 0s_{1/2}\,$ and (0.078 MeV, $3^-$) 
with $0f_{7/2}\to 0s_{1/2}\,$ for the ground-state doublet;  
(9.497 MeV, $5^+$) with $0f_{7/2}\to 0p_{3/2}\,$ and (9.633 MeV, $4^+$) 
with $0f_{7/2}\to 0p_{3/2}$ and $0p_{1/2}\,$ for the second main peak;
(18.181 MeV, $6^-$) with $0f_{7/2}\to 0d_{5/2}\,$ and  
(18.313 MeV, $5^-$) with $0f_{7/2}\to 0d_{3/2}$ and $0d_{5/2}\,$ 
for the third peak; 
and (24.619 MeV, $7^+$) with $0f_{7/2}\to 0f_{7/2}\,$ and   
(24.687 MeV, $6^+$) with $0f_{7/2}\to 0f_{7/2}$ and $0f_{5/2}\,$ 
for the fourth main peak.   
Exceptions are the states: (11.942 MeV, $1^+$) with dominant transition 
$1s_{1/2}\to 0s_{1/2}\,$, 
(13.327 MeV, $2^+$) with $0d_{3/2}\to 0s_{1/2}\,$, 
(27.814 MeV, $4^-$) with $0d_{5/2}\to 0p_{3/2}\,$, and 
(20.008 MeV, $3^-$) with $0d_{5/2}\to 0p_{3/2}$ and $0p_{1/2}$. 
Moreover the part of the spectrum with $\Lambda$ in the $f$ orbit reveals 
quite different distribution of the production strength than in 
Fig. 7 of Ref.~\cite{NPA881}.
Also the strength in between the main peaks, based on the proton hole 
in $d^{-1}_{3/2}$, is missing in our result based on the TD$_\Lambda$ 
approach. A possible explanation is that we would need to use an approach 
beyond TD$_\Lambda$, which includes coupling to the nuclear core 
excitation, in order to obtain this strength. An example of such an 
approach is EMPM$_\Lambda$, see Ref.~\cite{structure} for a more 
detailed discussion.   

In Figure~\ref{52Cr-YNG-SLA} one can also clearly observe shifts 
of the main peaks in the NF spectrum to larger excitation energies, 
which  depend on the $\Lambda$ orbital momentum. These shifts can be 
attributed to different strength of the $\Lambda$ spin-orbital parts 
of the NF and JA interactions.

The $\Lambda$ s-wave binding energies ($B_\Lambda$) for 
$^{52}_{~\Lambda}$V obtained in our approach are 20.7 and 18.6 MeV 
for the NF and JA interactions, respectively. 
These values are smaller by 1.1 and 3.2 MeV in comparison with 
the experimental value 21.8 $\pm$ 0.3 MeV presented in Table IV 
of the review paper~\cite{GHM2016}.  
Similarly to the case of $^{40}_{~\Lambda}$K and $^{48}_{~\Lambda}$K 
the JA interaction predicts smaller $B_\Lambda$ than NF suggesting 
a milder depth of the $\Lambda$--core-nucleus potential.   

%
%
\begin{figure}[htb]
\begin{center}
\includegraphics[width=6.5cm,angle=270]{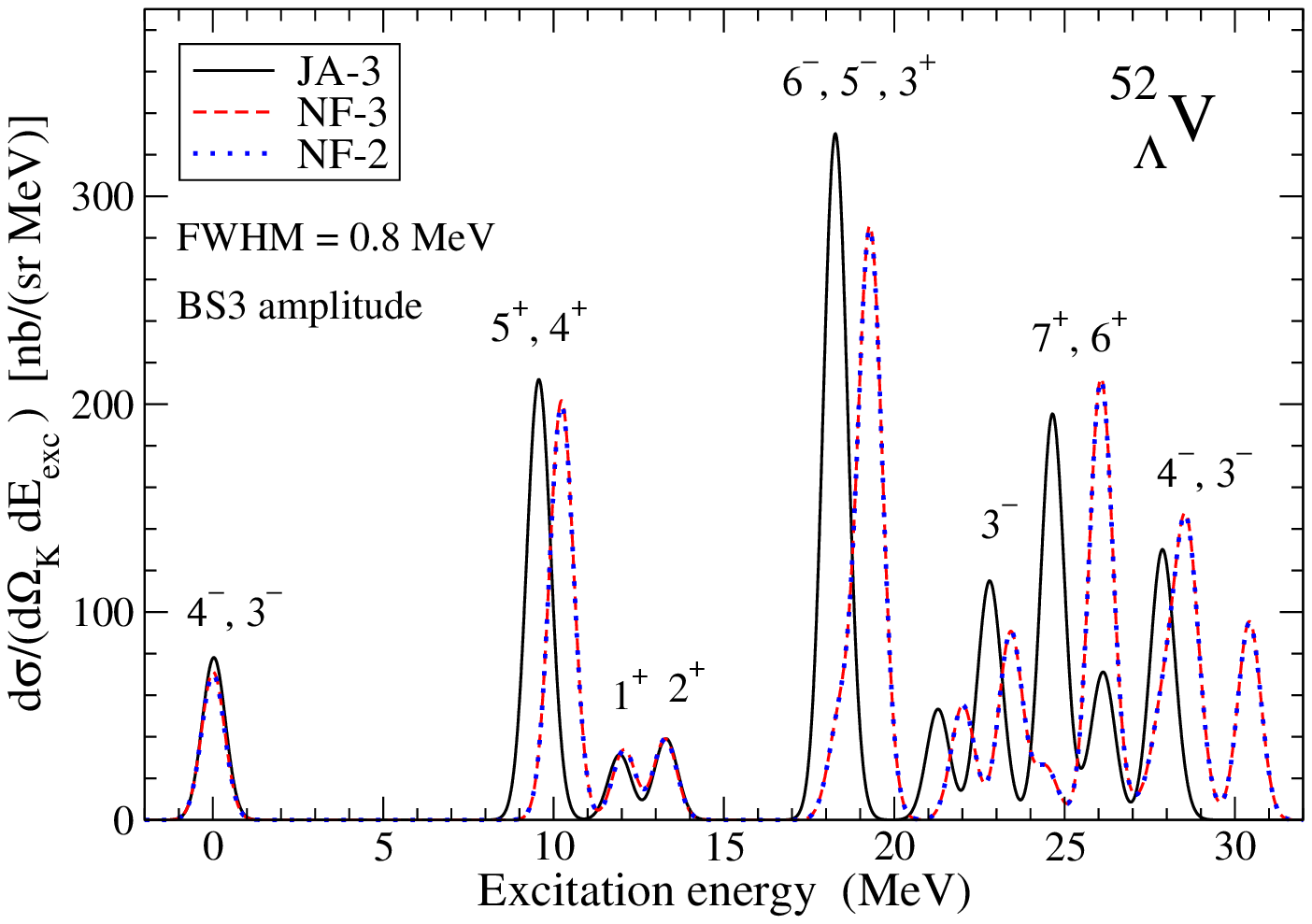}
\end{center}
\caption{The same as in Fig.~\ref{52Cr-YNG-SLA} but with the BS3 amplitude 
in the on-shell approximation (JA-3 and NF-3). The result NF-3 is compared 
with previous ``optimum on-shell approximation'' (NF-2).}
\label{52Cr-YNG-BS3}
\end{figure}
In Figure~\ref{52Cr-YNG-BS3} we compare the results calculated with the 
BS3 amplitude in the on-shell approximation for the JA and NF YNG 
interactions. The only difference of shapes of spectra in 
Figs.~\ref{52Cr-YNG-SLA} and \ref{52Cr-YNG-BS3} for a given YNG interaction 
is in magnitudes of the peaks which depend on the elementary amplitude and 
kinematics (on-shell or frozen-proton approximation). 
Note that in the frozen-proton approximation the proton and $\Lambda$ 
momenta are 0 and 362 MeV/c whereas in the on-shell approximation they are 
181 and 105 MeV/c (for the ground state), respectively. This difference 
might seem small but it does play a role, as shown in Figs.~\ref{208Pb-adep} 
and \ref{208Pb-edep} and in Ref.~\cite{fermi}. 
In the case of NF we also compare the new on-shell result (NF-3) with 
the previous optimum on-shell approximation (NF-2) used before 
in Refs.~\cite{fermi} and \cite{structure}. 
As we have already mentioned the difference between these spectra is 
almost negligible. This is because, although the magnitudes of the cross 
sections for separate hypernucleus states differ by about 1\%, the sign 
of the differences for various states in a multiplet is scattered on 
both sides (positive/negative) and therefore the overall shape of the 
peaks composed from several states remains almost unchanged. 
The small difference of the cross sections can be attributed to a small 
difference between the proton momentum $p_{opt}$ with 
$\cos\theta_{\Delta p}= -1$ and $p_{mean}$ for the proton single-particle 
wave function of the dominant transition, see Sec.~III.A. 

The results in Fig.~\ref{52Cr-YNG-BS3} demonstrate a measure of 
sensitivity of the spectra to using different forms of the effective 
YNG interactions in the DWIA calculations which is comparable to the 
width of the peaks (FWHM = 800 keV). We therefore presume that 
analysis of a good quality experimental spectrum obtained in the 
reaction $^{52}$Cr($e,e^\prime K^+$)$^{52}_{~\Lambda}$V  
could help in a more precise determination of 
the spin-dependent part of the YNG interaction, 
e.g., the strength of the $\Lambda$ spin orbital term.   
%
%
\subsection{$^{208}_{\;\;\;\Lambda}$Tl}
In this subsection we discuss results for the reaction 
$^{208}$Pb$(e,e'K^+)^{208}_{\;\;\;\Lambda}$Tl where both 
$^{208}$Pb and $^{208}_{\;\;\;\Lambda}$Tl are very heavy systems with 126 
neutrons which act as spectators in the impulse approximation still 
affecting kinematics of the reaction. The mass of the ground state of 
$^{208}$Pb is taken to be  $M_{Pb}^N=193.68769$ GeV and that of 
$^{208}_{\;\;\;\Lambda}$Tl is estimated to be 
$M_{Tl}^H= M_{Pb}^N-m_p +m_\Lambda +\varepsilon_p -\varepsilon_\Lambda 
\doteq 193.8482$ GeV considering that the difference of the binding  
energies is $m_\Lambda -m_p +\varepsilon_p -\varepsilon_\Lambda\approx 
160.5$ MeV. Determining precise values of the masses is quite important, 
remind the discussion of sensitivity of the cross sections to the 
hypernucleus mass in Sec.~III. 

The cross sections of $^{208}_{\;\;\;\Lambda}$Tl are calculated in 
the HF+TD$_\Lambda$ approach using the $NN$ interaction D16+DDT with 
$C_{\rho} = 4000$ MeV$\cdot$fm$^6$ and various YNG interactions and 
elementary amplitudes. The aim is to show a variety of predictions 
for the planned experiment E12-20-013 at JLab~\cite{E12-20-013}.  
The results are for the new kinematics in Hall C~\cite{Hall C}, 
$E_i= 2.24$ GeV, $E_f=0.74$ GeV, $\theta_e=8^\circ$, 
$\theta_{Ke}=11^\circ$, and $\Phi_K=180^\circ$, which gives 
$E_\gamma= 1.5$ GeV, $|\vec{p}_\gamma|= 1.511$ GeV/c, 
$Q^2= 0.0323$ (GeV/c)$^2$, $\theta_{\gamma e}= 3.9^\circ$, 
$\varepsilon = 0.591$, $\Gamma= 0.0140$ sr$^{-1}$GeV$^{-1}$, 
$\theta_{K\gamma}= 7.1^\circ$, and, in the case 
of the ground state, $|\vec{P}_K|_{mb}= 1.245$ GeV/c, 
$|\vec{\Delta}|= 315$ MeV/c, $\theta_{\Delta\gamma}= 29.2^\circ$, 
$|\vec{p}_{opt}|= 205$ MeV/c, and $|\vec{p}_\Lambda|= 110$ MeV/c. 
The kinematics is in the laboratory frame. 

\begin{figure}[htb]
\begin{center}
\includegraphics[width=6.9cm,angle=270]{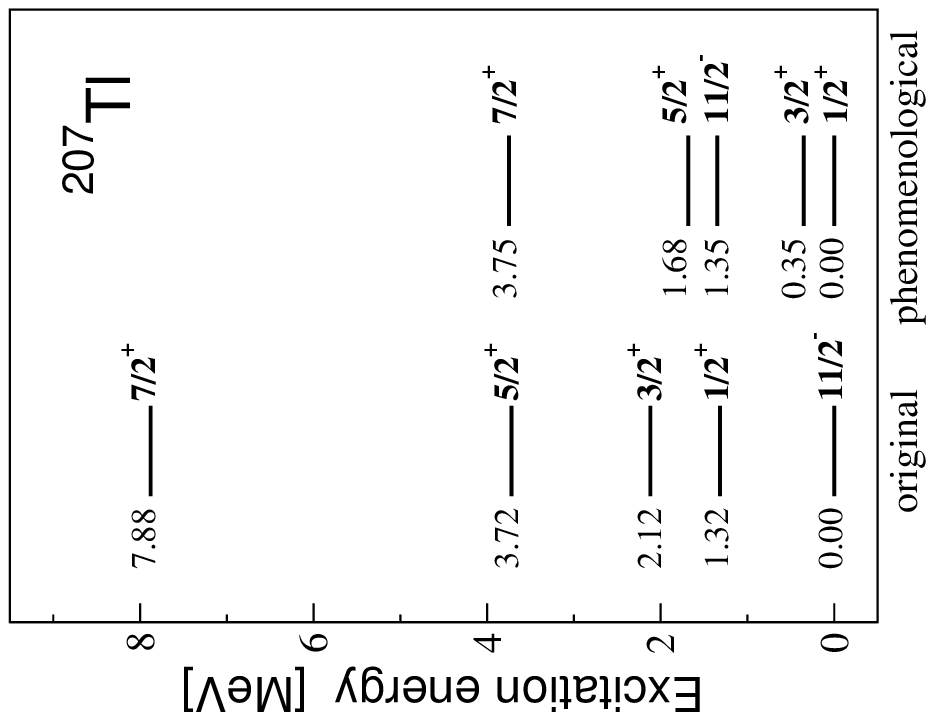}
\end{center}
\caption{The spectrum of the core nucleus $^{207}$Tl calculated by  
the HF method using the D16+DDT $NN$ interaction with $C_{\rho}$ = 4000 
MeV$\cdot$fm$^6$ and denoted as ``original'' in the left column is 
compared with phenomenological values obtained in 
experiments~\cite{Wilson,DataSheets,WittH} 
(the right column). The energies are in MeV.}
\label{spectrum-207Tl}
\end{figure}

The ground state of the target nucleus $^{208}$Pb is $J_{\sf A}^{P_{\sf A}}=0^+$ 
and its wave function is represented by the HF Slater determinant. We assume 
that $^{208}_{\;\;\;\Lambda}$Tl can be understood as $\Lambda$ bound in the core 
nucleus $^{207}$Tl which is a spectator in the impulse approximation. Therefore, 
it is important to describe well also the core nucleus in our approach. Within 
the HF method, the energy spectrum of $^{207}$Tl can be estimated from the proton 
single-particle energies $\varepsilon_{p}$. If we set the last occupied proton 
level (i.e. 0h$_{11/2}$) to 0.0 MeV we can take the relative energies of the 
occupied levels 2s$_{1/2}$, 1d$_{3/2}$, 1d$_{5/2}$, 0g$_{7/2}$ with 
respect to 0h$_{11/2}$ as an estimate of the low-lying energy spectrum of 
$\frac{11}{2}^-$, $\frac{1}{2}^+$, $\frac{3}{2}^+$, $\frac{5}{2}^+$, 
$\frac{7}{2}^+$ states in $^{207}$Tl. Such a spectrum of the core nucleus obtained 
within our HF calculation is shown in the left column in Fig.~\ref{spectrum-207Tl} 
and denoted as ``original'' because it was obtained with the original form 
of the $NN$ interaction. This original  spectrum is compared with 
the phenomenological one in the right column which was obtained from 
experiments~\cite{Wilson,DataSheets,WittH}. 
Note that the fifth single-particle state $\frac{7}{2}^+$ was observed 
in the $^{208}$Pb($\gamma,p$) and $^{208}$Pb($e,e^\prime p$) 
reactions~\cite{DataSheets,WittH} and was also considered 
in the calculations by Motoba and Millener~\cite{E12-20-013}. 
However, this state was not seen in the collision of two $^{208}$Pb 
nuclei~\cite{Wilson}. Instead the state $\frac{17}{2}^+$ at 3.813 MeV was 
observed.
%
%

The comparison demonstrates that the empirical spectrum is quite 
significantly squeezed down with respect to the theoretical (original) 
result and the ground-state in the theoretical spectrum $\frac{11}{2}^-$ 
lies above the state $\frac{3}{2}^+$ in the phenomenological spectrum. 
The phenomenological spectrum reflects that, in fact, the effective 
$NN$ interaction should be modified. In this study, we mimic such an effect 
by changing the proton 2s$_{1/2}$, 1d$_{3/2}$, 1d$_{5/2}$, 0g$_{7/2}$ 
single-particle energies $\varepsilon_p$ in order to reproduce the 
phenomenological $\frac{1}{2}^+$, $\frac{3}{2}^+$, $\frac{11}{2}^-$, 
$\frac{5}{2}^+$, and $\frac{7}{2}^+$ energies of $^{207}$Tl. These values 
of $\varepsilon_p$ enter the TD$_{\Lambda}$ calculation (see Eq. (11) 
in \cite{structure}) and, therefore, modify the OBDME.  
 
\begin{figure}[htb]
\begin{center}
\includegraphics[width=6.2cm,angle=270]{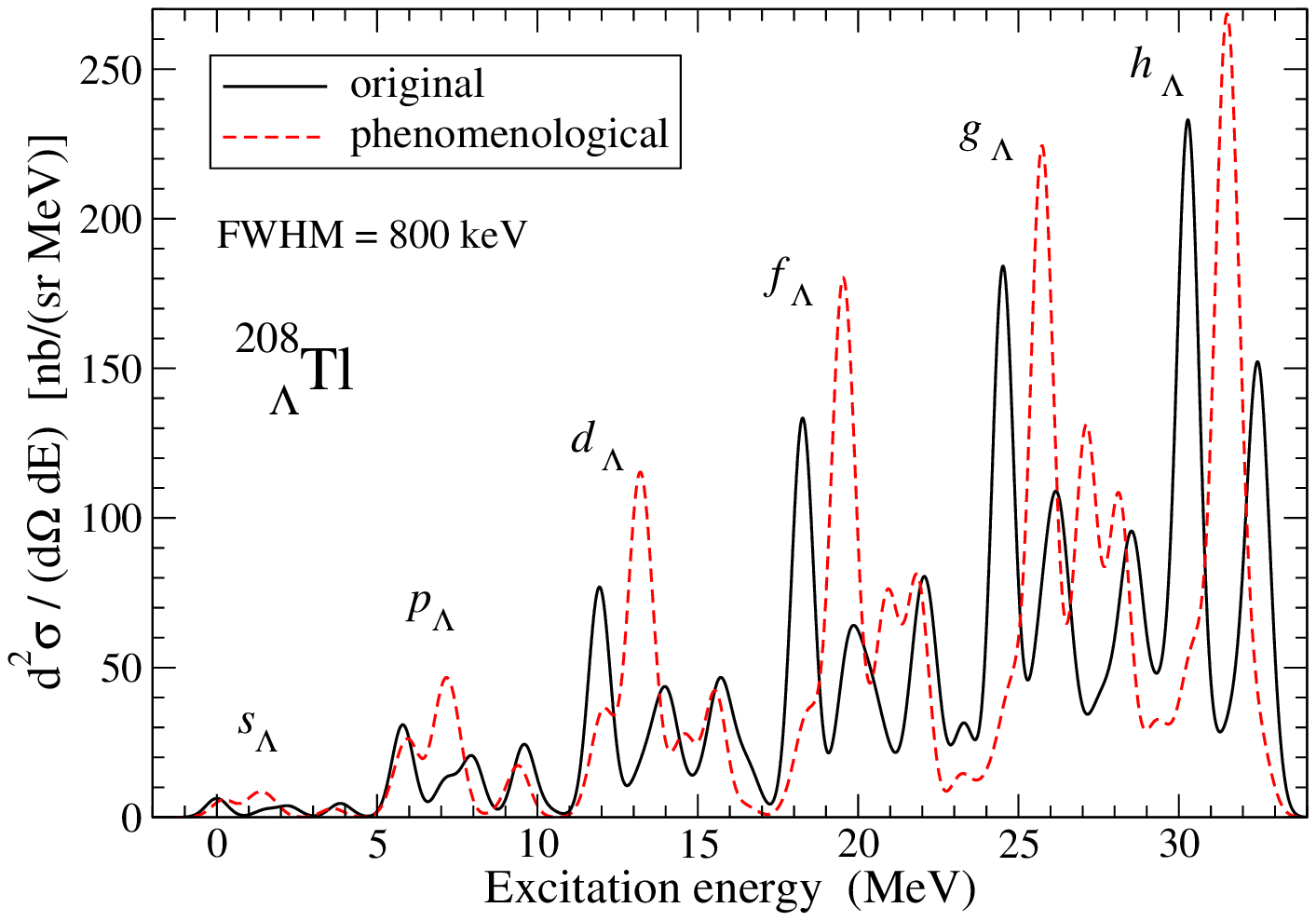}
\end{center}
\caption{The excitation spectrum in 
$^{208}$Pb$(e,e^\prime K^+)^{208}_{~~\Lambda}$Tl obtained consistently 
with the original and phenomenological spectra of 
$^{207}$Tl shown in Fig.~\ref{spectrum-207Tl}.}
\label{spectrum-L208Tlphen}
\end{figure}

The excitation spectra in electroproduction of $^{208}_{~~\Lambda}$Tl 
calculated with the original and phenomenological spectra of $^{207}$Tl 
are shown in Fig.~\ref{spectrum-L208Tlphen}. Both results were obtained 
using the BS3 amplitude in the on-shell approximation and the Nijmegen F 
YNG interaction with $k_{\sf F}= 1.34$ fm$^{-1}$. The results are for 
kinematics of the experiment E12-20-013 in Hall C.  
One can see that changing the spectrum of $^{207}$Tl results in 
a systematical shift of positions of the main peaks to larger energies. 
In general, the shape of the new excitation spectrum has changed with 
the phenomenological single-particle energies suggesting that ordering 
of some hypernucleus states has changed as well as the production 
cross sections.  
Note that assuming the phenomenological spectrum of $^{207}$Tl means 
a modification of the cross sections via the values of OBDME.  
%
%

The strength of the proton($\alpha$) $\to$ $\Lambda (\alpha^\prime)$ 
transition is given by the reduced matrix element of the one-body 
transition operator (OBDME) 
$(\Phi_{\sf H}||\left[b_{\alpha^\prime}^+\otimes a_\alpha\right]^J|| 
\Phi_{\sf A})$ and the radial integral 
${\cal R}^{LM}_{\alpha^\prime\alpha}$ where the sum in 
Eq.~(\ref{amplitude-3}) runs over a model space given by 
$\alpha\equiv (n\,l\,\,j)_{\sf p}$ and 
$\alpha^\prime\equiv(n^\prime l^\prime j^\prime)_\Lambda$. 
In case of heavy hypernuclei the model space is big and its dimension 
is restricted by a considered accuracy. We compared the results for 
$^{208}_{~~\Lambda}$Tl calculated only with the transitions satisfying 
$|{\rm OBDME}|\le 0.001$ and $|{\rm OBDME}|\le 0.0001$ and found tiny 
differences. Therefore we use the former restriction in our calculations. 
Moreover, in the sum typically one up to five transitions dominate, and 
therefore one can say that the given hypernucleus state is populated by 
a particular transition $\alpha\to\alpha^\prime$ with $J$, $L$, and $S$.  
Note that the following triangle conditions must be fulfilled for the 
quantum numbers: $\Delta(l^\prime,l,\,L)$, $\Delta(J_{\sf H},J_{\sf A},J)$, 
$\Delta(j^\prime,j,\,J)$, and $\Delta(J,\,L,\,S)$ with $J_{\sf A}=0$ and 
$S=0,1$ for the transitions without and with spin flip, respectively. 
Moreover, the orbital momenta must respect the parity of the final state, 
i.e., $P_{\sf H}\cdot P_{\sf A} = (-1)^L$, and $l+l^\prime +L$ must be an 
even number. The right single-particle quantum numbers to couple to the 
core nucleus must also satisfy  $\Delta(J_{\sf c},J_{\sf H},j^\prime)$, 
$\Delta(J_{\sf c},J_{\sf A},j)$, $P_{\sf c}\cdot P_{\sf A} = (-1)^l$, and 
$P_{\sf c}\cdot P_{\sf H} = (-1)^{l^\prime}$. 

The dominant transitions for selected hypernucleus states with the 
corresponding quantum numbers are shown in Tables~\ref{dominant} and 
\ref{dominant2}. We also show the cross sections and the states in the 
``original'' and ``phenomenological'' spectra of the core $^{207}$Tl 
which can be used to build up the ground state of $^{208}$Pb and the 
states of $^{208}_{~~\Lambda}$Tl. In Table~\ref{dominant} we compare 
the first and second doublet based on the ground and first excited states 
of the core nucleus. In the case of the original spectrum the first 
hypernucleus doublet is based on the ground state $\frac{11}{2}^-$ 
of $^{207}$Tl and the second doublet on the excited state $\frac{1}{2}^+$ 
at 1.317 MeV. 
This excitation energy then gives the splitting of about 1.52 MeV of the 
doublets. An analogous situation is in the case of the phenomenological 
spectrum with the states $\frac{1}{2}^+$ and $\frac{3}{2}^+$ split only 
by 0.351 MeV giving the moderate doublet splitting $\approx0.28$ MeV.
%
%
\begin{widetext}
\begin{center}
\begin{table}[htb]
\begin{tabular}{cccccccccccccc}
\multicolumn{14}{c}{original spectrum of $^{207}$Tl}\\
\hline
\multicolumn{2}{c}{core} & \multicolumn{3}{c}{proton} & 
\multicolumn{3}{c}{$\Lambda$} &
 
\multicolumn{2}{c}{hypernucleus} & & & & cross section\\
$E_{\sf c}$ & $J_{\sf c}^{P_{\sf c}}$ & $n$ & $l$ & $j$ & $n^\prime$ & 
$l^\prime$ & $j^\prime$ & $E_{\sf H}$ & $J_{\sf H}^{P_{\sf H}}$ & $J$ & 
$L$ & $S$ & [nb/sr]\\
0.000 & $\frac{11}{2}^-\ \ $ & 0 & 5 & $\frac{11}{2}\ \ $ & 
0 & 0 & $\frac{1}{2}\ \ $ &  0.000 & $6^-$ & 6 & 5 & 1 & 
3.66\vspace{1mm}\\
0.000 & $\frac{11}{2}^-\ \ $ & 0 & 5 & $\frac{11}{2}\ \ $ & 
0 & 0 & $\frac{1}{2}\ \ $ &  0.003 & $5^-$ & 5 & 5 & 0,1 & 1.74
\vspace{1mm}\\ \hline
 1.317 &  $\frac{1}{2}^+\ \ $ & 2 & 0 & $\frac{1}{2}\ \ $ & 0 & 0 & 
 $\frac{1}{2}\ \ $ & 1.485 & $0^+$ & 0 & 0 & 0 & 0.03\vspace{1mm}\\
 1.317 &  $\frac{1}{2}^+\ \ $ & 2 & 0 & $\frac{1}{2}\ \ $ & 0 & 0 & 
 $\frac{1}{2}\ \ $ & 1.519 & $1^+$ & 1 & 0 & 1 & 2.04\vspace{1mm}\\
 \hline\vspace{1mm}\\
\multicolumn{14}{c}{phenomenological spectrum of $^{207}$Tl}\\
\hline
\multicolumn{2}{c}{core} & \multicolumn{3}{c}{proton} & 
\multicolumn{3}{c}{$\Lambda$} &
 
\multicolumn{2}{c}{hypernucleus} & & & & cross section\\
$E_{\sf c}$ & $J_{\sf c}^{P_{\sf c}}$ & $n$ & $l$ & $j$ & $n^\prime$ & 
$l^\prime$ & $j^\prime$ & $E_{\sf H}$ & $J_{\sf H}^{P_{\sf H}}$ & $J$ & 
$L$ & $S$ & [nb/sr]\\
0.000 & $\frac{1}{2}^+$ & 2 & 0 & $\frac{1}{2}$ & 0 & 0 & $\frac{1}{2}$ &  
0.000 & $0^+$ & 0 & 0 & 0 & 0.03\vspace{1mm}\\
0.000 & $\frac{1}{2}^+$ & 2 & 0 & $\frac{1}{2}$ & 0 & 0 & $\frac{1}{2}$ &  
0.034 & $1^+$ & 1 & 0 & 1 & 2.07\vspace{1mm}\\ 
\hline
0.351 &  $\frac{3}{2}^+$ & 1 & 2 & $\frac{3}{2}$ & 0 & 0 & $\frac{1}{2}$ & 
0.313 & $2^+$ & 2 & 2 & 0,1 & 2.08\vspace{1mm}\\
0.351 &  $\frac{3}{2}^+$ & 1 & 2 & $\frac{3}{2}$ & 0 & 0 & $\frac{1}{2}$ & 
0.319 &  $1^+$ & 1 & 2 & 1 & 1.16\vspace{1mm}\\
 \hline
\end{tabular}
\caption{Doublets of hypernuclear states are shown with the quantum numbers 
of dominant transitions $p\to\Lambda$, the cross sections, and the 
corresponding states of the core nucleus. The cases with the ``original'' 
and ``phenomenological'' spectra of the core $^{207}$Tl are compared.} 
\label{dominant}
\end{table}
\end{center}
\end{widetext}

In Table~\ref{dominant} we can also observe that the hypernucleus state 
with the larger spin in the doublet possesses the larger cross section 
in accord with the rule that the state with the maximum spin in the 
multiplet is populated most strongly. Another well known fact is that 
at small kaon angles, considered in electroproduction of hypernuclei, 
the elementary production is dominated by its spin-flip ($S=1$) part. 
Therefore, the hypernucleus states with mere $S=0$ are only weakly 
populated whereas those with $S=1$ have significant cross sections. 
This dominance of the spin-flip amplitude is clearly seen in the second 
and first doublets of the upper and lower parts of the table, respectively.
This dominance also makes the difference of the ground-state cross sections  
for the original and phenomenological spectra, where the former is 
dominantly populated by the core-nucleus state $\frac{11}{2}^-$ allowing 
$S=1$, but the latter is given by the state $\frac{1}{2}^+$ requiring $S=0$.

In Table \ref{dominant2} we show the multiplets with $\Lambda$ in the $p$ 
and $d$ orbits, see also Fig.~\ref{spectrum-L208Tlphen}. 
It is apparent again that the maximum spin states 7$^+$ and 8$^-$ dominate 
the $p_\Lambda$ and $d_\Lambda$ multiplets, respectively, in both 
computational approaches. All transitions shown in Table~\ref{dominant2} 
are based on the proton hole in the $0h_{11/2}$ orbit which corresponds 
to the ground state of the core nucleus in the approach with the original 
spectrum but to the excited state at 1.348 MeV in the phenomenological 
spectrum. The difference 1.348 MeV then makes the shift of the main 
$p_\Lambda$ and $d_\Lambda$ peaks observed in 
Fig.~\ref{spectrum-L208Tlphen}. 
Recall that due to the selection rule the longitudinal parts of the cross 
section obtain significant contributions from the spin-flip amplitude 
($S=1$) only for the states $7^+$, $8^-$, and $6^-$ whereas the other 
states $6^+$, $7^-$, and $5^-$ receive their leading contributions only 
from the non spin-flip part of the elementary amplitude which is weaker in 
the studied kinematical region. This is also one of the reasons why 
the cross sections for the states $7^+$ and $8^-$ are so big. 
%
%
\begin{widetext}
\begin{center}
\begin{table}[htb]
\begin{tabular}{cccccccccccccc}
\multicolumn{14}{c}{original spectrum of $^{207}$Tl}\\
\hline
\multicolumn{2}{c}{core} & \multicolumn{3}{c}{proton} & 
\multicolumn{3}{c}{$\Lambda$} &
 
\multicolumn{2}{c}{hypernucleus} & & & & cross section\\
$E_{\sf c}$ & $J_{\sf c}^{P_{\sf c}}$ & $n$ & $l$ & $j$ & $n^\prime$ & 
$l^\prime$ & $j^\prime$ & $E_{\sf H}$ & $J_{\sf H}^{P_{\sf H}}$ & $J$ & 
$L$ & $S$ & [nb/sr]\\
0.000 & $\frac{11}{2}^-$ & 0 & 5 & $\frac{11}{2}$ & 
0 & 1 & $\frac{3}{2}$ &  5.804 & $7^+$ & 7 & 6 & 1 & 17.82\vspace{1mm}\\
0.000 & $\frac{11}{2}^-$ & 0 & 5 & $\frac{11}{2}$ & 
0 & 1 & $\frac{1}{2}$,$\frac{3}{2}$ &  5.794 & $6^+$ & 6 & 6 & 0,1 & 7.88\vspace
{1mm}\\ 
\hline
 0.000 &  $\frac{11}{2}^-$ & 0 & 5 & $\frac{11}{2}$ & 0 & 2 & 
 $\frac{3}{2}$, $\frac{5}{2}$ & 11.875 & $7^-$ & 7 & 7 & 0,1 & 
 18.02\vspace{1mm}\\
 0.000 &  $\frac{11}{2}^-$ & 0 & 5 & $\frac{11}{2}$ & 0 & 2 & 
 $\frac{3}{2}$, $\frac{5}{2}$ & 11.934 & $5^-$ & 5 & 5 & 0,1 & 
 0.20\vspace{1mm}\\
 0.000 &  $\frac{11}{2}^-$ & 0 & 5 & $\frac{11}{2}$ & 0 & 2 & 
 $\frac{3}{2}$, $\frac{5}{2}$ & 11.946 & $6^-$ & 6 & 5,7 & 1 & 
 1.27\vspace{1mm}\\ 
 0.000 &  $\frac{11}{2}^-$ & 0 & 5 & $\frac{11}{2}$ & 0 & 2 & 
 $\frac{5}{2}$ & 11.964 & $8^-$ & 8  & 7 & 1 & 41.78\vspace{1mm}\\
 \hline\vspace{1mm}\\
\multicolumn{14}{c}{phenomenological spectrum of $^{207}$Tl}\\
\hline
\multicolumn{2}{c}{core} & \multicolumn{3}{c}{proton} & 
\multicolumn{3}{c}{$\Lambda$} & \multicolumn{2}{c}{hypernucleus} & & & & 
cross section\\
$E_{\sf c}$ & $J_{\sf c}^{P_{\sf c}}$ & $n$ & $l$ & $j$ & $n^\prime$ & 
$l^\prime$ & $j^\prime$ & $E_{\sf H}$ & $J_{\sf H}^{P_{\sf H}}$ & $J$ & 
$L$ & $S$ & [nb/sr]\\
1.348 & $\frac{11}{2}^-$ & 0 & 5 & $\frac{11}{2}$ & 0 & 1 & $\frac{1}{2}$, 
$\frac{3}{2}$ &  6.974 & $6^+$ & 6 & 6 & 0,1 & 7.44\vspace{1mm}\\
1.348  & $\frac{11}{2}^-$ & 0 & 5 & $\frac{11}{2}$ & 0 & 1 & $\frac{3}{2}$ & 
6.984 & $7^+$ & 7 & 6 & 1 & 16.62\vspace{1mm}\\
\hline   
1.348  & $\frac{11}{2}^-$ & 0 & 5 & $\frac{11}{2}$ & 0 & 2 & 
$\frac{3}{2}$,$\frac{5}{2}$ & 13.109 & $7^-$ & 7 & 7 & 0,1 & 
10.65\vspace{1mm}\\
1.348  & $\frac{11}{2}^-$ & 0 & 5 & $\frac{11}{2}$ & 0 & 2 & $\frac{3}{2}$,
$\frac{5}{2}$ & 13.110 & $5^-$ & 5 & 5 & 0,1 & 0.63\vspace{1mm}\\
1.348  & $\frac{11}{2}^-$ & 0 & 5 & $\frac{11}{2}$ & 0 & 2 & $\frac{3}{2}$,
$\frac{5}{2}$ & 13.122 & $6^-$ & 6 & 5,7 & 1 & 1.37\vspace{1mm}\\
1.348  & $\frac{11}{2}^-$ & 0 & 5 & $\frac{11}{2}$ & 0 & 2 & $\frac{5}{2}$ 
& 13.144 & $8^-$ & 8 & 7 & 1 & 39.94\vspace{1mm}\\
 \hline
\end{tabular}
\caption{The same as in Table~\ref{dominant} but for the $\Lambda$ in the 
$p$ ($l^\prime=1$) and $d$ ($l^\prime=2$) orbits.} 
\label{dominant2}
\end{table} 
\end{center}
\end{widetext}

Predictions of excitation spectra in electroproduction of 
$^{208}_{~~\Lambda}$Tl are also compared with the previous result by 
Motoba and Millener utilized in the proposal of the Jlab experiment 
E12-20-013~\cite{E12-20-013}. This older calculation was done for 
photoproduction at $E_\gamma= 1.5$ GeV and $\theta_{K\gamma}= 0.5^\circ$ 
using the Saclay-Lyon amplitude A (SLA) in the frozen-proton approximation 
($p_{eff}=0$). The $\Lambda$ was assumed to be weakly coupled to the 
proton-hole states of $^{207}$Tl, which are strongly populated in the 
($e,e^\prime p$) and (d, $^3$He) reactions on $^{208}$Pb, and the $\Lambda$ 
single-particle energies were calculated from the Woods-Saxon potential.  
Our calculation for electroproduction was performed in the HF + TD$_\Lambda$ 
approach using the $NN$ interaction D16+DDT with 
$C_{\rho}$ = 4000 MeV$\cdot$fm$^6$ and the SLA amplitude in the 
frozen-proton approximation assuming kinematics $E_i= 4$ GeV, $E_f= 2.5$ 
GeV, $\theta_e= 2^\circ$, $\theta_{Ke}= 3.83^\circ$, and 
$\Phi_K= 180^\circ$, giving photoproduction kinematics close to that by 
Motoba and Millener, $E_\gamma = 1.5$ GeV, $|\vec{q}\,|= 1.504$ GeV/c, 
$Q^2= 0.012$ (GeV/c)$^2$, and $\theta_{K\gamma}= 0.5^\circ$. 
The relatively small value of $Q^2$ allows a comparison with 
the photoproduction calculation. 
%
%
\begin{figure}[htb]
\begin{center}
\includegraphics[width=6.2cm,angle=270]{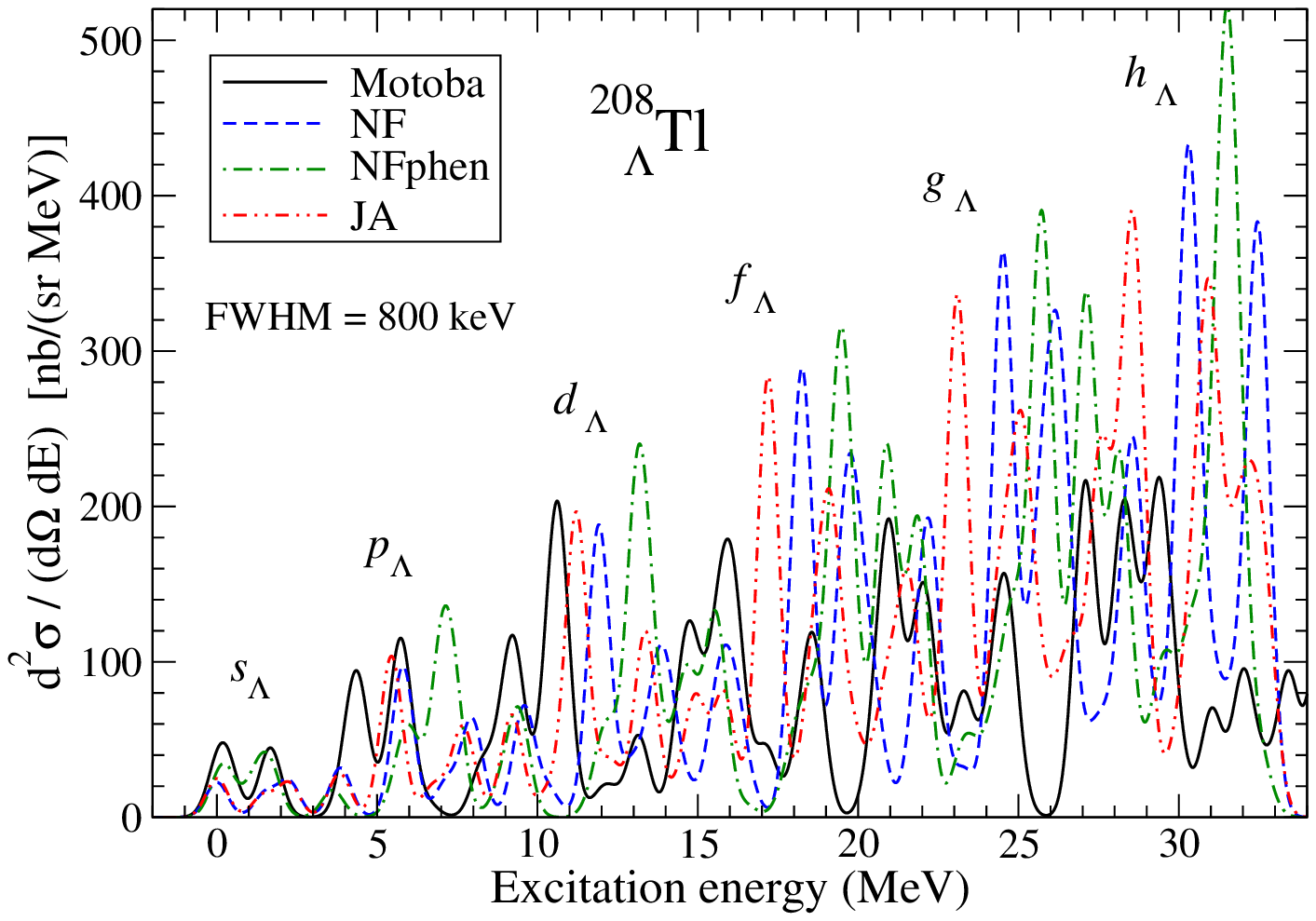}
\end{center}
\caption{Comparison of the result by Motoba and Millener~\cite{E12-20-013} 
with our calculations in the HF + TD$_\Lambda$ approach using the $NN$ 
interaction D16+DDT with $C_{\rho}$ = 4000 MeV$\cdot$fm$^6$ and the SLA 
amplitude in the frozen-proton approximation. The results were calculated 
with the Nijmegen F and J\"ulich A (JA) YNG interactions where the former 
was used with the original (NF) and phenomenological (NFphen) spectra  
of $^{207}$Tl in Fig.~\ref{spectrum-207Tl}. 
The curves are plotted with the width 800 keV for all states.}
\label{M-M}
\end{figure}

The results obtained with the Nijmegen F (NF) and J\"ulich A (JA) YNG 
interactions are compared with the result by Motoba and Millener in 
Fig.~\ref{M-M}. In the case of Nijmegen YNG interaction we show both 
variants of the calculation, with the original (NF) and phenomenological 
(NFphen) spectra of $^{207}$Tl (see Fig.~\ref{spectrum-207Tl}).  
One can see that in the new results the peaks with $\Lambda$ in the $d$, 
$f$, $g$, and $h$ (for NFphen also $p$) orbits are shifted to larger values 
of the excitation energy with respect to the Motoba-Millener result, where 
the shift is larger for the Nijmegen interaction. 
Moreover, the calculated cross sections for the high-laying states are 
systematically larger than the older values but they are a bit smaller 
for the $s_\Lambda$ states. The best agreement is achieved with the NF 
and JA interactions for the second $p_\Lambda$ peak, composed mainly of 
the very close states (5.792 MeV, 6$^+$), (5.804 MeV, 7$^+$), and 
(5.831 MeV, 5$^+$) with the NF and (5.425 MeV, 7$^+$), 
(5.456 MeV, 6$^+$), and (5.463 MeV, 5$^+$) with the JA interaction. 
These states are dominantly populated by the $0h_{11/2}\to 0p_{3/2}$ 
and $0h_{11/2}\to 0p_{1/2}$ transitions. 

In the case of the $s_\Lambda$ peaks in Fig.~\ref{M-M}, the NFphen result 
agrees quite well with that by Motoba and Millener. This is because the 
phenomenological spectrum of $^{207}$Tl is consistent with the spectrum 
used by Motoba and Milenner~\cite{E12-20-013}. Therefore, the left 
$s_\Lambda$ peak is based on the proton holes $2s^{-1}_{1/2}$ and 
$1d^{-1}_{3/2}$ and the right peak on $1d^{-1}_{5/2}$ and $0h^{-1}_{11/2}$,  
where the contribution of the latter is negligible in the Motoba-Millener 
results but is quite important in the NFphen result. 
In contrast, in the NF and JA results the left peak is given by the  
$0h^{-1}_{11/2}$ and $2s^{-1}_{1/2}$ holes whereas the right one 
is given by the $1d^{-1}_{3/2}$ hole.

Note that using the BS3 amplitude in the on-shell approximation changes 
mainly magnitudes of the main peaks at small excitation energies but at high 
energies the shape of the excitation spectrum is also changed, which is due 
to the large density of hypernuclear states populated with different 
strengths for the SLA and BS3 amplitudes.

One of the noticeable differences between our result and the Motoba-Millener 
result is presence of a rising background in our result, which causes the peaks 
in the high-energy region of the spectrum to systematically rise. 
This phenomenon can be attributed to a high density of relatively weakly 
populated states in our calculations. In the JA calculation, there are about 
630 states in the energy region 0--33 MeV, where most of the states are 
located above 10 MeV, and only 67 states are produced with cross section 
larger than 5 nb/sr.  On the other hand the older result includes only about 
160 states. Note that in the Motoba-Millener result presented in 
Ref.~\cite{E12-20-013} the background is partially simulated assuming 
a larger width of the peaks above the quasi-free threshold but in the 
result presented here we plotted the spectrum with the uniform width 800 keV.

\begin{figure}[htb]
\begin{center}
\includegraphics[width=6.2cm,angle=270]{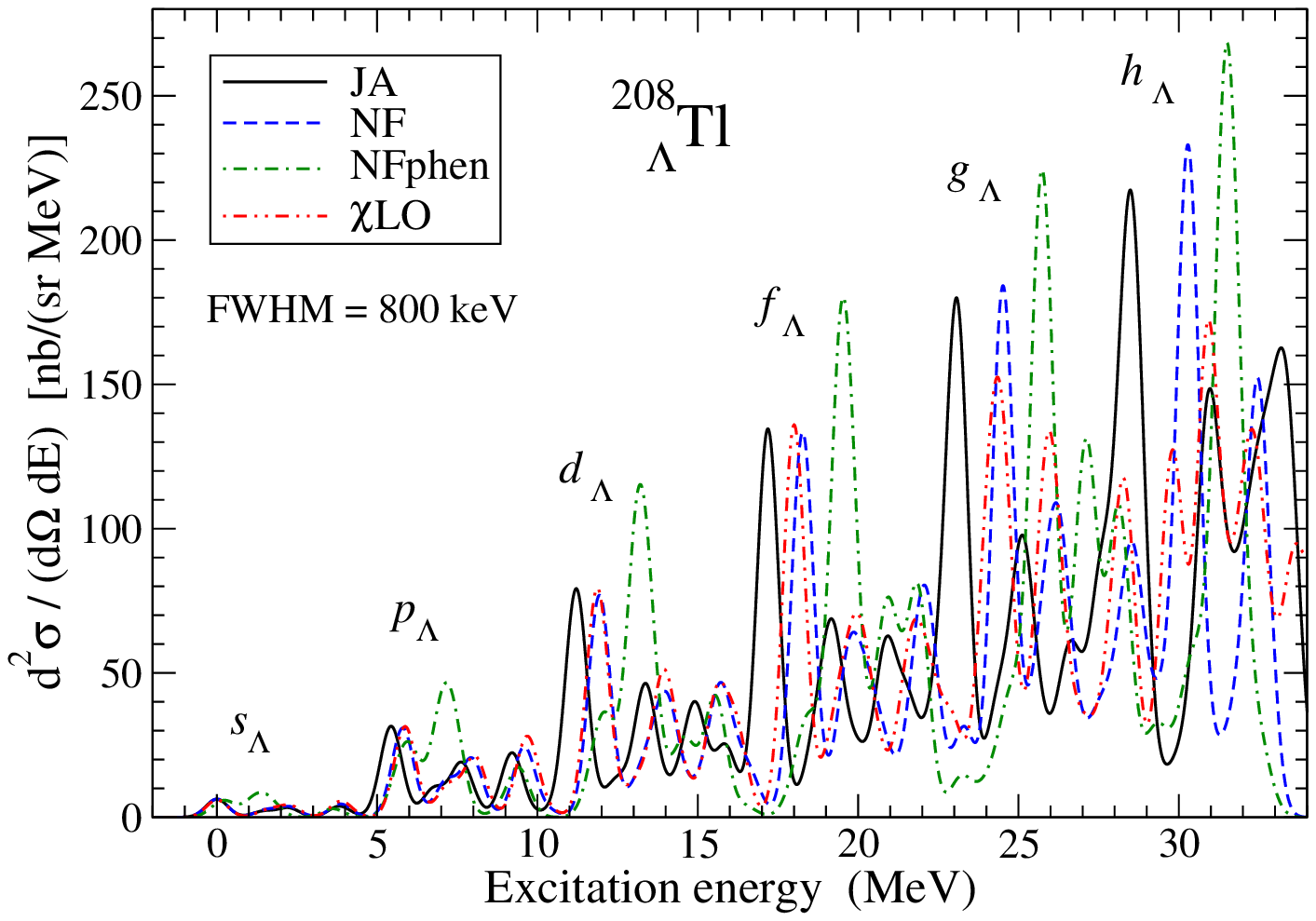}
\end{center}
\caption{Predicted spectrum of $^{208}_{~~\Lambda}$Tl in the Hall C 
kinematics of E12-20-013 calculated with the BS3 amplitude in the on-shell 
approximation for various forms of the baryon-baryon interaction: 
J\"ulich A (JA), Nijmegen F with original (NF) spectrum of $^{207}$Tl, 
Nijmegen F with phenomenological (NFphen) spectrum of $^{207}$Tl, and 
chiral leading order ($\chi$LO). For the G-matrix of the $\chi$LO   
potential, Eqs.~(\ref{Bethe}) and (\ref{EAV}) with 
$E_{\rm{av}} = -20.75$ MeV were used.}
\label{expE12}
\end{figure}

Finally, in Figure~\ref{expE12} we show predictions for the planned 
experiment E12-20-013 obtained with various forms of the effective $YN$  
interaction. The DWIA calculations for the Hall C kinematics~\cite{Hall C} 
were performed in the HF + TD$_\Lambda$ approach using the Daejeon 16 
$NN$ potential with the phenomenological DD term 
($C_{\rho}=4000$ MeV$\cdot$fm$^6$) and the BS3 amplitude in the on-shell 
approximation. The results are for the YNG interactions J\"ulich A (JA), 
chiral in the leading order ($\chi$LO), and Nijmegen F. The latter was 
considered with the original (NF) and phenomenological (NFphen) spectra  
of the core nucleus $^{207}$Tl. The G-matrix for the $\chi$LO potential  
was calculated with $E_{\rm{av}} = -20.75$ MeV in Eqs.~(\ref{Bethe}) 
and (\ref{EAV}). 

One can see that even if the nature of the NF and $\chi$LO YNG 
interactions is different the results for the excitation spectrum 
in Fig.~\ref{expE12} do not differ too much. In contrast, 
positions of the main $p_\Lambda$--$h_\Lambda$ peaks predicted 
by NF are shifted to larger excitation energies with respect 
to the peaks from JA where the magnitude of the shift 
systematically rises with increasing orbital momentum of $\Lambda$. 
This, similarly to the case of $^{52}_{~\Lambda}$V, suggests 
a different strength of the $\Lambda$ spin-orbital part of these YNG 
interactions. Note that the $NN$ interaction is the same for NF, JA, 
and $\chi$LO but it differs for NFphen, which results in a more 
pronounced change of the excitation spectrum. 
The main peaks from NFphen are shifted to larger excitation 
energies and the corresponding production cross sections are larger 
than those for the original NF. These effects can be attributed to 
the modification of the $NN$ interaction due to using the 
phenomenological values of the excitation energies of $^{207}$Tl 
given in Fig.~\ref{spectrum-207Tl}. 

The $\Lambda$ s-wave binding energies ($B_\Lambda$) for 
$^{208}_{~~\Lambda}$Tl predicted in this work are 27.44, 27.27, 
24.46, and 28.39 MeV for the YNG interactions NF, NFphen, JA, 
and $\chi$LO, respectively. The results for NF and NFphen are 
quite close to the experimental value 26.9 $\pm$ 0.8 MeV obtained 
in the ($\pi^+$,K$^+$) reaction as given in Table IV 
of Ref.~\cite{GHM2016}. Note that improving spectrum of the core 
nucleus $^{207}$Tl in NFphen leads to a better agreement with the 
data on $B_\Lambda$. 
Similarly to the previous cases JA gives a smaller $B_\Lambda$ 
but $\chi$LO predicts a larger $B_\Lambda$ than the experimental value.

The results in Fig.~\ref{expE12} show variety of predictions coming 
from different forms of the YNG interactions. Good quality 
experimental data, allowing determination of positions of the main 
peaks in the spectrum, could prefer one of the considered YNG 
interactions. Moreover, the measured magnitudes of the peaks related 
to averaged cross sections in the multiplet can shed more light 
on dynamics of 
the process, the elementary amplitude, and the kaon distortion. 
%
%
%
%
\section{Summary and conclusions}
We studied various effects in DWIA calculations of the cross sections 
in electroproduction of hypernuclei beyond the $p$ shell. We analyzed 
the results in view of kaon distortion, kinematical effects, and 
various forms of the effective hyperon-nucleon interaction. 
The latter was aimed especially at illustrating differences of the 
predictions of the reaction cross sections due to a variety of possible 
forms of the effective hyperon-nucleon interactions. 

We showed that the damping effect of the kaon re-scattering is rising 
with the hypernucleus mass, amounting 40--60\% for $^{40}_{~\Lambda}$K 
and 50--80\% for $^{208}_{~\Lambda}$Tl where the states with deeply bound 
$\Lambda$ are more affected. The kaon distortion, accounted for via the 
eikonal approximation with the first-order optical potential, depends 
mainly on behavior of the nucleon density in the peripheral region. 

We have found that the momenta and the cross sections apparently depend  
on a tiny modification of the hypernucleus mass due to excitation energy, 
denoted as the hypernucleus-mass shift. This mass shift modifies the 
proton, kaon, and $\Lambda$ momenta as well as the momentum transfer by 
a few percent, which changes values of the elementary amplitude and 
radial integrals. These changes then suppress the cross section by 
several percent, which is important mainly for the highly excited 
states.  

We have elaborated our previously suggested optimum on-shell 
approximation~\cite{fermi}, specifying the value of the angle 
$\theta_{\Delta p}$ which was a free parameter before. 
In the new variant we suggest equating the magnitude of the proton effective 
momentum to a mean momentum of the proton in the single-particle orbit, 
which corresponds to the dominant transition given by the OBDME. The angle 
is then calculated from the energy conservation in 
the elementary vertex. This we denote as ``on-shell approximation'',  
where the elementary amplitude is on-shell with a fully specified 
(in the coplanar kinematics) non zero value of the proton momentum.

We demonstrated the influence of various values of the proton effective 
momentum on the cross sections for electroproduction of 
$^{208}_{~~\Lambda}$Tl. In the given kinematics, differences between 
the proton- and $\Lambda$-frozen approximations amount up to 30\% in 
the photon energy region 1.2--2.2 GeV. Similarly to the $p$-shell 
hypernuclei~\cite{fermi}, the effects are more apparent at small kaon 
angles. Influence of the selection rule~\cite{fermi}, which controls 
contributions from the longitudinal mode of the virtual photon to 
the reduced amplitude, was also discussed for selected states 
of $^{208}_{~~\Lambda}$Tl.  

We updated and discussed spectra in electroproduction of 
$^{40}_{~\Lambda}$K and $^{48}_{~\Lambda}$K using a new effective $NN$  
interaction, Daejeon 16 $NN$ interaction complemented with 
the phenomenological DD term, and two forms of the YNG interaction, 
Nijmegen F (NF) and J\"ulich A (JA). Moreover, the optical potential 
used in accounting for the kaon distortion was calculated with the 
HF nucleon density, which is a more realistic description than 
using the HO density in our previous analysis~\cite{structure}.
The results were calculated in kinematics of the planned experiment 
E12-15-008 at Jefferson Lab and they extend the discussion of 
predictions for the experiment initiated in Ref.~\cite{structure}. 

The obtained spectra reveal clearly separated main peaks which 
do not differ too much for the NF and JA YNG interactions. 
However, the calculated $\Lambda$ $s$-wave binding energies 
($B_\Lambda$) accord well with the other analysis by Friedman 
and Gall~\cite{FG} only for the NF interaction. 
The $B_\Lambda$ predicted by JA for $^{40}_{~\Lambda}$K and 
$^{48}_{~\Lambda}$K are smaller almost by 2 MeV than those in 
Ref.~\cite{FG}, suggesting that JA is weaker than NF in our formalism. 

We also discussed predictions for the electroproduction spectrum of 
$^{52}_{~\Lambda}$V. This spectrum clearly shows separated multiplets 
with $\Lambda$ in the $s$, $p$, and $d$ orbits which might be 
suitable for an experimental investigation.  

In the case of the $^{208}$Pb target we discussed the results 
in view of various forms of the baryon-baryon interaction. 
We showed predictions with the J\"ulich A, Nijmegen F, and 
chiral leading-order ($\chi$LO) YNG interactions and compared 
them with older calculations by Motoba and Millener utilized 
in preparing the experimental proposal E12-20-013 at Jefferson Lab. 
Despite their different natures, the NF and $\chi$LO YNG interactions 
predict similar production spectrum for $^{208}_{~~\Lambda}$Tl 
in kinematics of E12-20-013. On the other hand, the NF and JA 
interactions reveal a different strength of the $\Lambda$ spin orbital 
part, shifting the main $p_\Lambda - h_\Lambda$ peaks for NF to 
higher energies with rising magnitude. 

We also modified our calculations by using a phenomenological 
spectrum of the core nucleus $^{207}$Tl and, together with NF, 
we utilized this modified description in predicting the excitation 
spectra for $^{208}_{~~\Lambda}$Tl in kinematics of the experiment 
E12-20-013. This modified calculation with the NF YNG interaction 
gives more realistic values of $B_\Lambda$ that the other 
calculations with JA and $\chi$LO.

In conclusion, let us note that in this analysis we used other NN 
interaction and more forms of the YNG interactions than in our 
previous work~\cite{structure} to discuss predictions for heavy 
hypernuclei such as $^{208}_{~~\Lambda}$Tl. 
One of the goals of the analysis is to show sensitivity 
of the excitation spectra to using various interactions. 
Particularly, we aim at demonstrating if the new data from the planned 
JLab experiment with the resolution of about 800 KeV are able to 
select a realistic YNG interaction among the available interactions. 
%
%
\section*{ACKNOWLEDGEMENT}
The work was supported by the Czech Science Foundation GACR, 
Grant No. P203-24-10180S. Computational resources were provided by 
the e-INFRA CZ project (ID90254), supported by the Ministry of Education, 
Youth and Sports of the Czech Republic. 
%
%
%


\begin{thebibliography}{99}
\bibitem{Hall C} Kinematics of the experiment E12-20-013 was changed 
      with respect to that in Hall A~\cite{E12-20-013}. The new kinematics 
      in Hall C is $E_i= 2.24$ GeV, $E_f= 0.74$ GeV, $\theta_e= 8^\circ$, 
      $\theta_{Ke}= 11^\circ$, and $\Phi_K= 180^\circ$.  
      Private communication by S.N. Nakamura.
\end{thebibliography}
\end{document}